\documentclass{article}

\pdfoutput=1
\usepackage{arxiv}

\usepackage[utf8]{inputenc} 
\usepackage[T1]{fontenc}    
\usepackage[hidelinks]{hyperref}       
\usepackage{url}            
\usepackage{booktabs}       
\usepackage{amsfonts}       
\usepackage{nicefrac}       
\usepackage{microtype}      
\usepackage{lipsum}
\usepackage{mathtools}
\usepackage{graphicx}
\graphicspath{ {./images/} }
\usepackage{natbib}
\usepackage{amsmath}
\usepackage{makecell}
\usepackage{adjustbox}
\usepackage[table]{xcolor}
\usepackage{enumitem}

\usepackage{multirow}
\usepackage[title]{appendix}
\usepackage[flushleft]{threeparttable}

\def\bSig\mathbf{\Sigma}

\usepackage{natbib}
\usepackage{amsmath}
\usepackage{mathtools}
\usepackage{makecell}
\usepackage{adjustbox}
\usepackage[table]{xcolor}
\usepackage{siunitx}
\usepackage[mathscr]{euscript}
\usepackage{amsfonts}
\usepackage{xcolor}
\usepackage{xr}

\newcommand{\Ind}{\mathrel{\text{\scalebox{1.07}{$\perp\mkern-10mu\perp$}}}}

\title{Double Robust Variance Estimation with Parametric Working Models}

\author{Bonnie E. Shook-Sa$^{1,*}$, 
Paul N. Zivich$^{2}$, Chanhwa Lee$^{1}$, Keyi Xue$^{1}$,  
\\ \textbf{Rachael K. Ross$^{3}$, Jessie K. Edwards$^{2}$, Jeffrey S. A. Stringer$^{4}$, and Stephen R. Cole$^{2}$} \\
\\
{$^{1}$Department of Biostatistics, University of North Carolina at Chapel Hill, Chapel Hill, North Carolina, U.S.A.} \\
{$^{2}$Department of Epidemiology, University of North Carolina at Chapel Hill, U.S.A.} \\
{$^{3}$Department of Epidemiology, Mailman School of Public Health, Columbia University, New York City, New York, U.S.A.} \\
{$^{4}$Department of Obstetrics and Gynecology, University of North Carolina at Chapel Hill, 
Chapel Hill, North Carolina, U.S.A.} \\ 
\small{$*$}\footnotesize{bshooksa@email.unc.edu}}

\begin{document}
 \maketitle
 \begin{abstract}
Doubly robust estimators have gained popularity in the field of causal inference due to their ability to provide consistent point estimates when either an outcome or exposure model is correctly specified. However, for nonrandomized exposures the influence function based variance estimator frequently used with doubly robust estimators of the average causal effect is only consistent when both working models (i.e., outcome and exposure models) are correctly specified. Here, the empirical sandwich variance estimator and the nonparametric bootstrap are demonstrated to be doubly robust variance estimators. That is, they are expected to provide valid estimates of the variance leading to nominal confidence interval coverage when only one working model is correctly specified. Simulation studies illustrate the properties of the influence function based, empirical sandwich, and nonparametric bootstrap variance estimators in the setting where parametric working models are assumed. Estimators are applied to data from the Improving Pregnancy Outcomes with Progesterone (IPOP) study to estimate the effect of maternal anemia on birth weight among women with HIV.
	\end{abstract}

	
\section{Introduction}\label{sec:Intro}
Two common approaches for estimating causal effects of point exposures with observational study data are inverse probability of treatment weighting (IPTW) and g-computation. IPTW estimators rely on estimation of the probability of exposure given a set of covariates, i.e., the propensity score. Alternatively, g-computation is a direct standardization approach where the outcome is modeled as a function of the exposure and covariates. Both approaches provide consistent estimators of the average causal effect (ACE) if the corresponding model is correctly specified, but these estimators are generally not equivalent due to differing parametric constraints imposed by modeling. As the functional forms of these working, or nuisance, models are typically unknown, correct specification remains an obstacle to valid effect estimation.

Augmented inverse probability weighted (AIPW) estimators and targeted maximum likelihood estimators (TMLE) combine estimates from the working models in such a way that resulting estimators are “doubly robust.” That is, they are consistent when either the exposure or outcome model is correctly specified, but not necessarily both  \citep{robins2007comment,FunkDR,kang2007demystifying,vanderLaanRubin2006}. Two common variations of AIPW estimators are an estimator that incorporates estimated propensity scores from the exposure model and pseudo-outcomes from the outcome model (i.e., the classic AIPW) and a g-computation type estimator where the outcome model is weighted by the IPTW (i.e., the weighted regression AIPW). TMLE is a variation of the weighted regression AIPW estimator that instead combines IPTW and the outcome model through a ``targeting'' model fit using the outcome predictions and weights. While primarily developed with machine learning approaches for estimation of nuisance parameters \citep{vanderLaanRubin2006}, TMLE can also be applied with parametric working models. 

Doubly robust estimators also offer precision advantages over IPTW estimators. When both working models are correctly specified, the three doubly robust estimators previously discussed are semiparametric efficient, having the smallest asymptotic variance of all IPTW estimators \citep{robins1994estimation,vanderLaanRubin2006,gabriel2023inverse}. Given these potential precision advantages and that the true specification of both models is typically unknown, doubly robust approaches have gained popularity.

The influence function based variance estimator commonly used with doubly robust methods has a major limitation. When estimating the average causal effect from observational data, this variance estimator is only consistent if both working models are correctly specified, i.e., this variance estimator is not doubly robust. This limitation has been highlighted previously \citep{FunkDR,daniel2014double}, yet the influence function based variance estimator is still commonly used. A systematic review published in October 2023 found that 67\% of papers published using TMLE did not indicate how variances were estimated. Of the 33\% that did report the method, the majority (72\%) relied on the influence function based variance estimator \citep{smith2023application}. Widespread application of the influence function based variance estimator is aided by available software, as it is used for confidence interval (CI) construction in the R \texttt{AIPW} and \texttt{tmle} packages and the \texttt{zEpid} Python package. 

When estimating the average causal effect, performance of the influence function based variance estimator depends on which working model is misspecified. When the outcome model is correctly specified but the propensity model is misspecified, the influence function based variance estimator is inconsistent \citep{munoz2012population}. In this setting, consistent point estimators may be paired with CIs that do not achieve the nominal level of coverage. If the propensity model is correctly specified but the outcome model is misspecified, the influence function based variance estimator is conservative \citep{munoz2012population}. Precision advantages of doubly robust estimators may be lost with application of a conservative variance estimator. These properties of the influence function based variance estimator do not necessarily apply for other estimands and in the randomized setting (see Section \ref{sec:discussion}).  

Alternative, but less commonly used, variance estimation approaches are doubly robust, as they are expected to provide valid estimates of the variance leading to nominal CI coverage when only one working model is correctly specified. Two such approaches are the nonparametric bootstrap and the empirical sandwich variance estimator based on M-estimation theory. The doubly robust property of the nonparametric bootstrap has been illustrated previously \citep{FunkDR}. M-estimation has been proposed for use with doubly robust estimators \citep{Lunceford2004,daniel2014double}, but we are not aware of prior work highlighting the doubly robust property of the empirical sandwich variance estimator. The M-estimation approach is equivalent to using the correct influence function with correction terms for estimation of nuisance parameters \citep{Stefanski2002}.

The recent systematic review \citep{smith2023application} did not mention application of the empirical sandwich variance estimator in any reviewed papers. M-estimation is appropriate with finite-dimension parametric working models and is generally not compatible with many machine learning approaches. SuperLearner was used in the majority of studies considered in the systematic review (62\%), and the empirical sandwich variance estimator would be inappropriate for these studies. However, logistic regression was used in 20\% of studies considered, and an additional 18\% of studies did not indicate what methods were used for nuisance parameter estimation. The lack of use of the empirical sandwich variance estimator in these settings suggests that the doubly robust property of the empirical sandwich variance estimator is not well known. Among published studies that reported the variance estimation approach, 21\% applied the bootstrap. However, knowledge of its doubly robust property may or may not be widespread.  

We aim to estimate the effect of maternal anemia on birth weight among women with HIV. Prior research has shown links between maternal anemia and low birth weight \citep{azizah2022effect}, and women with HIV experience anemia more frequently than women without HIV \citep{levine2001prevalence}. This effect is estimated with data from approximately 800 participants from the Improving Pregnancy Outcomes with Progesterone (IPOP) study. IPOP was a randomized controlled trial conducted in Lusaka, Zambia to evaluate the effectiveness of a therapeutic for prevention of preterm birth and stillbirth among women with HIV. Additional details about the IPOP trial have been published previously \citep{price2021weekly}. Covariates collected during the trial allow for examination of the relationship between anemia and birth weight. Doubly robust estimators may be preferred over IPTW or g-computation approaches due to uncertainty in working model specifications and potential efficiency advantages.

This paper demonstrates how M-estimation and the nonparametric bootstrap can be applied with AIPW estimators and TMLE to obtain doubly robust point \textit{and} variance estimators with parametric working models. In Section \ref{sec:methods}, three common doubly robust estimators are described along with three approaches for inference: the influence function based variance estimator, the empirical sandwich variance estimator, and the nonparametric bootstrap. Estimators are applied to IPOP data in Section \ref{sec:example}. In Section \ref{sec:sims}, the doubly robust properties of the empirical sandwich variance estimator and the nonparametric bootstrap are demonstrated in a simulation study and are contrasted with the performance of the influence function based variance estimator. The benefits and limitations of each variance estimation approach are discussed in Section \ref{sec:discussion}. 

\section{Methods}
\label{sec:methods}
\subsection{Preliminaries}
\label{sec:methods-preliminaries}
Consider an observational study that seeks to estimate the effect of a binary exposure $X$ on an outcome $Y$. Assume $n$ independent and identically distributed copies of $O_i=(X_i,Y_i,Z_i)$ are observed, where $Z$ are a set of baseline covariates. Let the potential outcomes for individual $i$ under exposure and no exposure be denoted by $Y_i^1$ and $Y_i^0$ such that by causal consistency $Y_i=X_iY_i^1+(1-X_i)Y_i^0$. Assume the set of covariates $Z$ satisfy $Y^x \Ind X \mid Z$ for $x \in \{0,1\}$, i.e., conditional exchangeability. Also assume positivity holds such that $\Pr(X=x \mid Z=z)>0$ for all $z$ such that $dF_Z(z)>0$ and $x \in \{0,1\}$, where $F_Z$ is the cumulative distribution function of $Z$. The estimand is the average causal effect ($ACE$), $\mu^1-\mu^0$ where $\mu^x=E(Y^x)$ for $x \in \{0,1\}$. As demonstrated elsewhere (e.g., \cite{Lunceford2004}), each causal mean $\mu^x$ is identifiable based on the observed data under the above assumptions. Specifically, $\mu^x=E\{E(Y \mid X=x, Z)\}$ and $\mu^x= E\{I(X=x)Y / \Pr(X=x \mid Z)\}$. The former expression motivates g-computation estimators, while the latter motivates IPTW estimators. Unless otherwise specified, vectors are assumed to be column vectors.

\subsection{Estimators}
\label{sec:methods-estimators}

Three doubly robust estimators of $ACE$ are considered, each of which combines estimates from the propensity and outcome models. The approach for modeling the propensity score is common across the methods, while there are slight differences in fitting the outcome model. In practice, the propensity score for each participant, $e_i=\Pr(X_i=1 \mid Z_i)$, is often estimated from the fitted model $\mbox{logit}(e_i)=G_{i}^T\alpha$, where $G_i=g(Z_i)$ is a vector of predictors for participant $i$ for some user-specified function $g$ of $Z_i$ and $\alpha$ is the vector of regression coefficients from the propensity model. The predicted propensity score for each participant is calculated as $\widehat{e}_i=e(Z_i,\widehat{\alpha})=\mbox{logit}^{-1}(G_{i}^T \hat{\alpha})$ where $\hat{\alpha}$ is the maximum likelihood estimate (MLE) of $\alpha$. The IPTW is then estimated for each participant as $\hat{W}_i=X_i\hat{e}_i^{-1}+(1-X_i)(1-\hat{e}_i)^{-1}$. 

\subsubsection{Classic AIPW estimator}
\label{sec:PIAIPW}
For the classic AIPW estimator, the $Y \mid Z,X$ model is fit and used to predict pseudo-outcomes for each participant under both exposures. That is, $a_x(Z_i,\hat{\gamma})=\hat{E}(Y_i \mid Z_i,X_i=x)$ for $x \in \{0,1\}$, where $\hat{\gamma}$ are the MLEs of the parameters from the assumed outcome model. Then, $ACE$ is estimated by
\begin{equation} \label{eq:pluginAIPW}
    \widehat{DR}_{C}= \hat{\mu}_{C}^1 - \hat{\mu}_{C}^0
\end{equation} where $\hat{\mu}_{C}^1=n^{-1}\sum_{i=1}^n  \hat{e}_i^{-1}\{X_iY_i-(X_i-\hat{e}_i)a_1(Z_i,\hat{\gamma})\}$ and $\hat{\mu}_{C}^0=n^{-1}\sum_{i=1}^n  (1-\hat{e}_i)^{-1}\{(1-X_i)Y_i+(X_i-\hat{e}_i)a_0(Z_i,\hat{\gamma})\}$. The estimator (\ref{eq:pluginAIPW}) was originally proposed by \cite{robins1994estimation} and was further examined in simulation studies, such as \cite{Lunceford2004}, \cite{kang2007demystifying}, and \cite{FunkDR}. 

\subsubsection{Weighted regression AIPW estimator}
\label{sec:wtdAIPW}
For the weighted regression AIPW estimator, the parameters of the $Y \mid Z,X$ model are estimated with IPTW-weighted maximum likelihood estimation. Pseudo-outcomes are obtained for each participant under both exposures, i.e., $b_x(Z_i,\hat{\beta})=\hat{E}(Y_i \mid Z_i,X_i=x)$ for $x \in \{0,1\}$, where $\hat{\beta}$ are the MLEs from the weighted outcome model. The weighted regression AIPW estimator is 
\begin{equation} \label{eq:wtdAIPW}
    \widehat{DR}_{WR}= \hat{\mu}_{WR}^1 - \hat{\mu}_{WR}^0
\end{equation} where $\hat{\mu}_{WR}^x=n^{-1}\sum_{i=1}^n b_x(Z_i,\hat{\beta}) $ for $x \in \{0, 1\}$. The estimator (\ref{eq:wtdAIPW}) has been evaluated previously and is expected to be more stable than (\ref{eq:pluginAIPW}) when IPTWs are extreme due to bounding within the parameter space \citep{kang2007demystifying,robins2007comment,Vansteelandt2012}.

\subsubsection{TMLE}
\label{sec:TMLE}
TMLE consists of a two-stage process. In the first stage, $Y \mid Z,X$ is modeled as in Section \ref{sec:PIAIPW}, and $a_x(Z_i,\hat{\gamma})$ are calculated for $x \in \{0,1\}$. In the second stage, referred to as the targeting stage, corrections to outcome regression predictions from stage one are made that incorporate the estimated propensity scores. Different methods have been proposed for fitting the targeting model; here, the weighted regression approach is considered, as it is thought to handle random positivity violations better \citep{van2011propensity}. First, assuming $Y$ is binary, the models $\mbox{logit}\{\Pr(Y_i=1 \mid Z_i, X_i=x)\}=\eta_x+\mbox{logit}\{a_x(Z_i,\hat{\gamma})\}$ are fit for $x \in \{0, 1\}$. Note the only parameter in this model is $\eta_x$, as $\mbox{logit}\{a_x(Z_i,\hat{\gamma})\}$ is a fixed offset. Parameters $\eta_x$ are estimated using weighted maximum likelihood, with weights $(1-X)\hat{W}_i$ and $X\hat{W}_i$ for the $x=0$ and $x=1$ models, respectively. The stage two pseudo-outcomes are computed as $c_x(O_i, \hat{\gamma}, \hat{\eta}_x)=\mbox{expit}[\mbox{logit}\{a_x(Z_i,\hat{\gamma})\}+\hat{\eta}_x]$. 
Then, the TMLE estimator of $ACE$ is 
\begin{equation} \label{eq:TMLE}
    \widehat{DR}_{TMLE}= \hat{\mu}_{TMLE}^1 - \hat{\mu}_{TMLE}^0
\end{equation} where $\hat{\mu}_{TMLE}^x=n^{-1}\sum_{i=1}^n c_x(O_i, \hat{\gamma}, \hat{\eta}_x)$ for $x \in \{0, 1\}$. For continuous outcomes, a logit model is still recommended for the targeting stage to ensure bounding within the parameter space \citep{gruber2010targeted,gruber2012tmle}. Here, the outcome is scaled prior to implementing TMLE, i.e., $Y_i^*=(Y_i-a)/(b-a)$ is defined, where $(a,b)$ are the bounds of $Y$. Then, $Y_i$ is replaced with $Y_i^*$ in the steps above, i.e.,  $E(Y_i^* \mid Z_i, X_i = x)$ is modeled with a logit link, and $c_x(O_i, \hat{\gamma}, \hat{\eta}_x)=\mbox{expit}[\mbox{logit}\{a_x(Z_i,\hat{\gamma})\}+\hat{\eta}_x](b-a)+a$ re-scales the outcome back to the original bounds of $Y$. 

TMLE methods were developed by \cite{vanderLaanRubin2006} and are frequently paired with machine learning techniques for nuisance modeling. When parametric models are used in both stages of estimation, (\ref{eq:TMLE}) is expected to be similar to (\ref{eq:wtdAIPW}) \citep{Tran2019}. 

\subsection{Variance Estimation}
\label{sec:Methods-var-est}
Variance estimation for doubly robust estimators (\ref{eq:pluginAIPW}), (\ref{eq:wtdAIPW}), and (\ref{eq:TMLE}) is not straightforward, because the variance of each estimator depends on estimating the parameters of the working models. Here, three methods for variance estimation are considered, each of which can be used to construct Wald-typed CIs for the $ACE$. 

\subsubsection{Influence function based variance estimator}
Estimators (\ref{eq:pluginAIPW}), (\ref{eq:wtdAIPW}), and (\ref{eq:TMLE}) are suggested by the efficient influence function \citep{luque2018targeted,gabriel2023inverse}. The commonly used influence function based variance estimator has the form $\hat{V}(\widehat{DR})_{IF}=n^{-2}\sum_{i=1}^n \hat{I}_i^2$, where 
$\hat{I}_i=\hat{e}_i^{-1}X_iY_i-(1-\hat{e}_i)^{-1}(1-X_i)Y_i-\{\hat{e}_i^{-1}(1-\hat{e}_i)^{-1}(X_i-\hat{e}_i)\}\{(1-\hat{e}_i)\hat{Y}_i^1+\hat{e}_i\hat{Y}_i^0\}-\widehat{DR}$ \citep{gruber2012tmle}. Here, pseudo-outcomes $\hat{Y}_i^x$ equal $a_x(Z_i,\hat{\gamma})$, $b_x(Z_i,\hat{\beta})$, and $c_x(O_i, \hat{\gamma}, \hat{\eta}_x)$ for the classic AIPW, weighted regression AIPW, and TMLE methods, respectively, and $\widehat{DR}$ equals (\ref{eq:pluginAIPW}), (\ref{eq:wtdAIPW}), and (\ref{eq:TMLE}) for the three corresponding estimators. 

The influence function based variance estimators $\hat{V}(\widehat{DR})_{IF}$ are consistent for the variance of (\ref{eq:pluginAIPW}), (\ref{eq:wtdAIPW}), and (\ref{eq:TMLE}) when both propensity score and outcome models are correctly specified \citep{daniel2014double,gruber2012tmle}. When one of the two working models is misspecified, the influence function based variance estimator is not consistent. As noted in the introduction, the estimator $\hat{V}(\widehat{DR})_{IF}$ is conservative when the weight model is correctly specified but the outcome model is misspecified, and may be conservative or anti-conservative when the outcome model is correctly specified but the weight model is misspecified. 

\subsubsection{Empirical sandwich variance estimator}
The estimators (\ref{eq:pluginAIPW}), (\ref{eq:wtdAIPW}), and (\ref{eq:TMLE}) can each be expressed as the solution to a set of unbiased estimating equations. The M-estimator for each set of estimating equations, $\hat{\theta}$, is the solution (for $\theta$) to $\sum_{i=1}^n \psi_q(O_i,\theta)=0$, where $\psi_q(O_i,\theta)$ are the estimating functions corresponding to each of the three estimators $q \in \{1,2,3\}$ with parameter vector $\theta$. Parameter vectors and estimating functions for each estimator are provided in Appendix A.

As further discussed in Appendix A, each set of estimating equations is unbiased when at least one of the working models is correctly specified. This result has been demonstrated previously for the classic and weighted regression AIPW estimators \citep{shook2024exposure,gabriel2023inverse}, and is demonstrated for TMLE in Appendix B. Because (\ref{eq:pluginAIPW}), (\ref{eq:wtdAIPW}), and (\ref{eq:TMLE}) are each solutions to unbiased estimating equation vectors, it follows under suitable regularity conditions \citep{Stefanski2002} that for each estimator, $\sqrt{n}(\hat{\theta}-\theta) \xrightarrow[]{d} N \left(0, V(\theta) \right)$. Here, $V(\theta)=A(\theta)^{-1}B(\theta)\{A(\theta)^{-1}\}^{T}$ where $A(\theta)=E\{-\partial \psi_q(O_i; \theta) / \partial \theta\}$ and $B(\theta)=E\{\psi_q(O_i; \theta)\psi_q(O_i; \theta)^{T}\}$. The quantities $A(\theta)$ and $B(\theta)$ can be consistently estimated by replacing each expected value with its empirical counterpart, resulting in the empirical sandwich variance estimator $\hat{V}(\hat{\theta})$. The variance of (\ref{eq:pluginAIPW}), (\ref{eq:wtdAIPW}), and (\ref{eq:TMLE}) can be consistently estimated by the bottom right element of its corresponding $\hat{V}(\hat{\theta})$, denoted as $\hat{V}(\widehat{DR})_{ES}$. Importantly, the estimating equations are unbiased when one or both working models are correctly specified, making the empirical sandwich variance estimator a doubly robust variance estimator.

\subsubsection{Nonparametric Bootstrap}
The nonparametric bootstrap can also be used to estimate the variance of (\ref{eq:pluginAIPW}), (\ref{eq:wtdAIPW}), and (\ref{eq:TMLE}). While variations exist, the nonparametric bootstrap typically involves selecting $B$ independent, with replacement resamples of size $n$ from the observed sample. The estimator of interest is applied to each resample $b$ to obtain $\widehat{DR}_b$, where $\widehat{DR}_b$ equals (\ref{eq:pluginAIPW}), (\ref{eq:wtdAIPW}), or (\ref{eq:TMLE}) applied to resample $b$. Then, $\hat{V}(\widehat{DR})_{NB}=\sum_{b=1}^B \{\widehat{DR}_b-\widehat{DR}^*\}^2 / (B-1)$, where $\widehat{DR}^*=B^{-1}\sum_{b=1}^B \widehat{DR}_b$. For large samples, the nonparametric bootstrap is expected to provide valid estimates of the variance for smooth functions of solutions to smooth estimating equations \citep{davison1997bootstrap}, including (\ref{eq:pluginAIPW}), (\ref{eq:wtdAIPW}), and (\ref{eq:TMLE}). Therefore, the nonparametric bootstrap is a doubly robust variance estimator. However, the bootstrap's performance may suffer in certain settings, e.g., in the presence of outliers \citep{davison1997bootstrap}.

\section{Example: Improving Pregnancy Outcomes with Progesterone}
\label{sec:example}

The doubly robust estimators described in Section \ref{sec:methods-estimators} were applied to IPOP data for estimation of the effect of maternal anemia on birth weight. The exposure, maternal anemia, was defined as having a baseline hemoglobin level below 10.5 g/dL. The IPOP data were limited to the 782 participants (98\%) with no prior preterm births and with measured birth weights and baseline hemoglobin values. Note 14 participants experienced stillbirth. Birth weights for these participants were included in the analysis; we return to this issue below. It was assumed that potential outcomes are independent of the exposure conditional on the covariates listed in Appendix Table A1. 

\begin{figure}
\begin{center}
  \includegraphics[width=1\columnwidth]{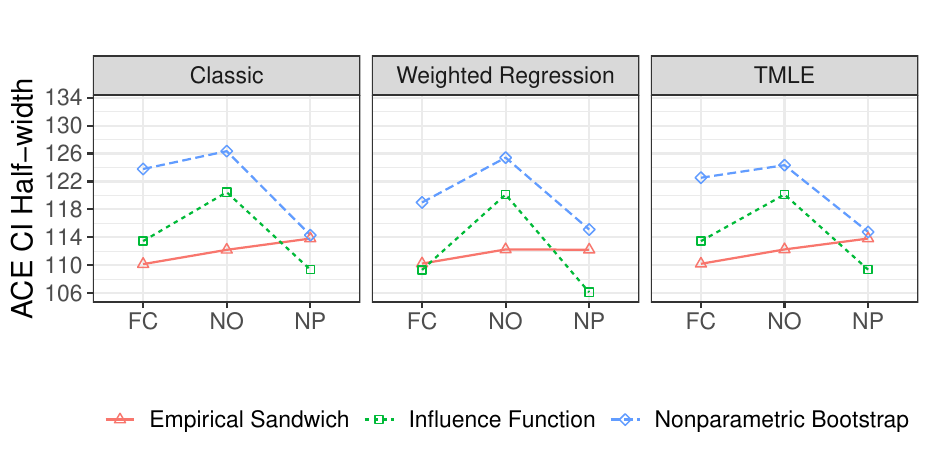}
  \caption{95\% confidence interval (CI) half-widths by model specification and estimator. FC=Full Covariate Set, NO=Naive Outcome Model, NP=Naive Propensity Model}
  \label{fig:halfwidths}
\end{center}
\end{figure}

To demonstrate differences in the variance estimators presented in Section \ref{sec:Methods-var-est}, each doubly robust estimator was applied with three sets of model specifications. First, the covariates in Appendix Table A1 were included in both propensity and outcome models. Then, a naive outcome model was fit that included only the exposure variable, while the propensity model included the full set of covariates. Under this specification, the outcome model was reasonably thought to be misspecified due to its simplicity. Finally, a naive propensity model was fit that included only an intercept term, i.e., assuming the propensity score is constant, while the outcome model included the full set of covariates. This propensity model was similarly thought to be misspecified. For all estimators, maternal height was modeled using restricted cubic splines with four knots placed at the 5th, 35th, 65th, and 95th percentiles. The empirical sandwich variance estimator was computed using \texttt{geex} in R \citep{saul2020calculus} and \texttt{delicatessen} in Python \citep{zivich2022delicatessen}. The bootstrap was based on 5000 resamples. Point estimates and corresponding 95\% CIs for each method and model specification are presented in Table \ref{tab:App_Results}, with 95\% CI half-widths compared in Figure \ref{fig:halfwidths}.

The doubly robust methods provided similar point estimates of the $ACE$ for all model specifications, with estimates of approximately -40g, though all 95\% CIs included zero. Precision estimates based on the empirical sandwich variance estimator and the influence function based variance estimator were similar when the full covariate set was included in both working models. Under the naive outcome model, standard errors for the influence function based variance estimator were larger than those for the empirical sandwich variance estimator. When the propensity model was naive, standard errors for the influence function based variance estimator were smaller than those for the empirical sandwich variance estimator. Note there was more fluctuation in 95\% CI half-widths across model specifications for the influence function based variance estimator than the empirical sandwich variance estimator (Figure \ref{fig:halfwidths}). Bootstrap standard errors were larger than those of the empirical sandwich and influence function based variance estimators. This may be due to the presence of extreme birth weights in the data (Appendix Figure A1).

There are notable limitations associated with this analysis. First, the $ACE$ is likely not the most informative estimand to address this research question. The $ACE$ contrasts average birth weight under settings where all women have anemia and no women have anemia. It is challenging to envision a setting where all women would be susceptible to anemia, so the estimators (\ref{eq:pluginAIPW}), (\ref{eq:wtdAIPW}), and (\ref{eq:TMLE}) may lack a meaningful causal interpretation. While these statistical quantities might still be beneficial, more informative estimands could also be considered to address this research question, including a comparison of $E(Y^0)$ with the natural course, $E(Y)$ \citep{hubbard2008population} or contrasts of stochastic intervention distributions \citep{kennedy2019nonparametric}. Additionally, as with all observational studies, there may be uncontrolled confounding not captured in the observed set of covariates. This concern may be heightened in settings where causal consistency is also questionable \citep{hernan2011compound}, i.e., there may be multiple interventions to modify anemia status. Finally, the analysis included participants who experienced stillbirth. Future work could reexamine this problem using alternative approaches for competing events.

\begin{table}
\caption{Estimated effect of maternal anemia on birth weight by model specification and estimator}
\label{tab:App_Results}
\centering

\begin{tabular}{l c c c c c c c} 
\hline
  & $\widehat{DR}$  & ES-SE & ES-95\% CI &   IF-SE & IF-95\% CI &   NB-SE & NB-95\% CI\\
\hline
\textbf{Full Covariate Set}	&		&		&		&		&		\\
Classic AIPW	&	-37	&	56	&	(-147, 73)	&	58	&	(-151, 76)	&	63	&	(-161, 86)	\\
Wtd regression AIPW	&	-41	&	56	&	(-151, 69)	&	56	&	(-151, 68)	&	61	&	(-160, 78)	\\
TMLE	&	-37	&	56	&	(-147, 73)	&	58	&	(-151, 76)	&	63	&	(-160, 85)	\\
\\

\textbf{Naive Outcome}	&		&		&		&		&		\\
Classic AIPW	&	-36	&	57	&	(-148, 76)	&	61	&	(-156, 84)	&	64	&	(-162, 90)	\\
Wtd regression AIPW	&	-36	&	57	&	(-148, 77)	&	61	&	(-156, 84)	&	64	&	(-161, 90)	\\
TMLE	&	-36	&	57	&	(-148, 77)	&	61	&	(-156, 84)	&	63	&	(-160, 89)	\\
\\

\textbf{Naive Propensity}	&		&		&		&		&		\\
Classic AIPW	&	-41	&	58	&	(-154, 73)	&	56	&	(-150, 69)	&	58	&	(-155, 74)	\\
Wtd regression AIPW	&	-47	&	57	&	(-159, 65)	&	54	&	(-153, 59)	&	59	&	(-162, 68)	\\
TMLE	&	-41	&	58	&	(-154, 73)	&	56	&	(-150, 69)	&	59	&	(-155, 74)	\\
\\

  \hline 
\end{tabular}
\begin{tablenotes}
      \item Note: SE=standard error, ES=empirical sandwich, IF=influence function, NB=nonparametric bootstrap, CI=confidence interval, Wtd=weighted. The naive outcome model included only the exposure, and the naive propensity model included only an intercept. The nonparametric bootstrap was based on 5000 resamples. 

    \end{tablenotes}
\end{table}

\section{Simulation Study}
\label{sec:sims}
A simulation study was conducted to compare the empirical properties of the influence function based variance estimator, the empirical sandwich variance estimator, and the nonparametric bootstrap in conjunction with each of the three doubly robust methods discussed in Section \ref{sec:methods}. Simulations were conducted with $n=800$, similar to the sample size in the example in Section \ref{sec:example}. 

\subsection{Simulation Setup}
\label{sec:sim-setup}
Three covariates ($Z_1$, $Z_2$, $Z_3$), the exposure ($X$), and potential outcomes ($Y^0$ and $Y^1$) were simulated. The covariate $Z_1$ was distributed Normal with mean $155$ and standard deviation $7.6$. Two binary covariates $Z_2$ and $Z_3$ were simulated from Bernoulli distributions with means $0.25$ and $0.75$, respectively. The exposure $X$ was simulated from a Bernoulli distribution with mean $\mbox{expit}(15-0.1 Z_1 + 2.5 Z_2 - Z_3 - 0.02 Z_1 Z_2 + 0.005 Z_1 Z_3)$. Potential outcomes $Y^1$ and $Y^0$ under exposure and no exposure, respectively, were simulated from a normal distribution with mean $E(Y^x)=1000+ 11.5 Z_1 + 100 Z_2 -15 Z_1 Z_2 +25x-5.5 x Z_1 -30 x Z_2 +5 x Z_1 Z_2$ and standard deviation $\sigma=400$ for $x \in \{0,1\}$. Under this data generating mechanism, the $ACE$ was approximately $-60$. The marginal distributions of the exposure and outcome were modeled after the example in Section \ref{sec:example}. To examine performance of the estimators under the null, simulations were also conducted where $E(Y^1)$ was equal to $E(Y^0)$, as defined above, such that $ACE=0$.

The estimators $\widehat{DR}_{C}$, $\widehat{DR}_{WR}$, and $\widehat{DR}_{TMLE}$ were applied to 5000 simulated samples. Estimators were computed four ways: (1) propensity and outcome models correctly specified, (2) propensity models correctly specified but outcome models misspecied, (3) outcome models correctly specified but propensity models misspecified, and (4) both models misspecified. Misspecified propensity models included only an intercept and a linear term for $(Z_1-155)^2$, and misspecified outcome models included only an intercept and linear terms for $X$ and $(Z_1-155)^2$. 

Each of the three variance estimators described in Section \ref{sec:Methods-var-est} was applied. For the nonparametric bootstrap, 1000 resamples were included in each iteration for estimation of the bootstrap standard error, excluding any resamples where working models failed to converge. Simulation results were summarized by empirical bias, average standard error (ASE), empirical standard error (ESE), standard error ratio (SER=ASE/ESE), and empirical 95\% CI coverage. The ratio of the variance estimate for each simulation and the ESE, i.e., the variance ratio, was also summarized. That is, $VR=se_s / ESE$, where $se_s=\sqrt{\hat{V}(\widehat{DR})}$ for a given doubly robust estimator, model specification, and variance estimator for simulation $s$.

\subsection{Simulation Results}
\label{sec:sim-results}
\begin{figure}
\begin{center}
  \includegraphics[width=\columnwidth]{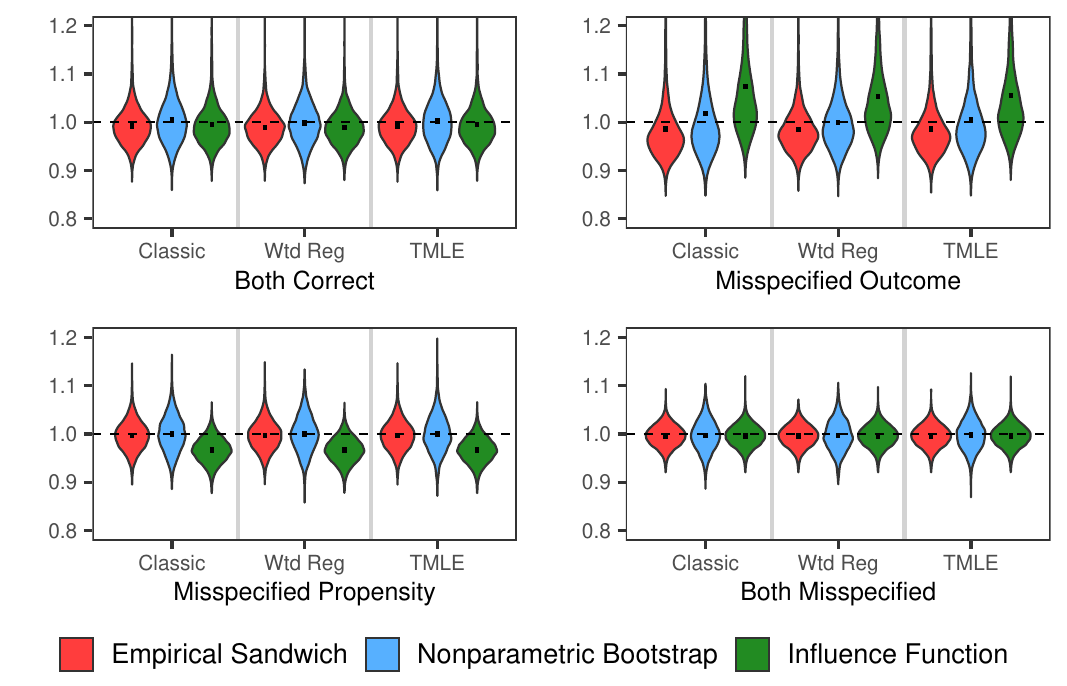}
  \caption{Ratio between each simulation's estimated standard error and the empirical standard error by estimator and model specification, continuous outcome, n=800, $\sigma=400$, 5000 simulations. $ACE$ was approximately -60. Black squares denote the mean variance ratio (=SER). Results exclude one simulation where models failed to converge. The $0.33\%$ of correct model specification simulations, $4.02\%$ of misspecified outcome model simulations, and $0.004\%$ of misspecified propensity model simulations where the ratio was above 1.2 or below 0.8 are not displayed.}
  \label{fig:SimresVardiff}
\end{center}
\end{figure}

\begin{figure}
\begin{center}
  \includegraphics[width=\columnwidth]{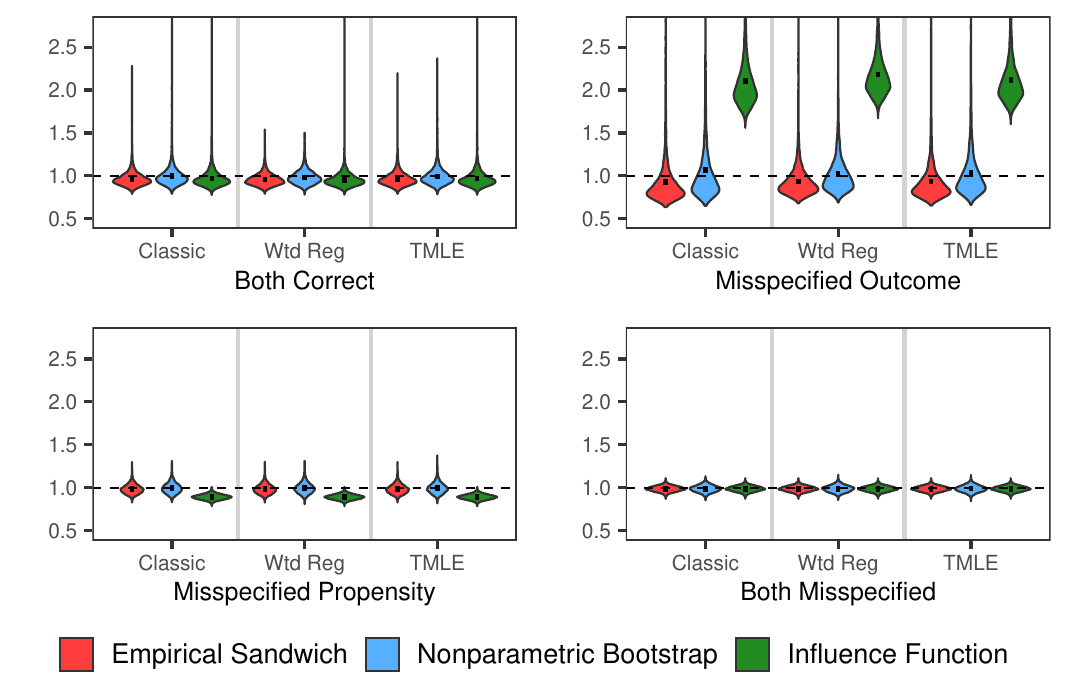}
  \caption{Ratio between each simulation's estimated standard error and the empirical standard error by estimator and model specification, continuous outcome, n=800, $\sigma=400$, 5000 simulations under the null. Black squares denote the mean variance ratio (=SER). The $0.01\%$ of correct model specification simulations and $1.72\%$ of misspecified outcome model simulations where the ratio was above 2.75 or below 0.5 are not displayed.}
  \label{fig:SimresVardiffNull}
\end{center}
\end{figure}

Both point and variance estimators performed as expected in simulations. The results of the simulation study are presented in Table \ref{tab:simres} and Figures \ref{fig:SimresVardiff} and \ref{fig:SimresVardiffNull}. When at least one model was correctly specified, the classic AIPW, weighted regression AIPW, and TMLE displayed minimal bias, but all were substantially biased when both working models were misspecified. Under correct specification of both working models, all three variance estimators tracked closely with the ESE, resulting in SERs close to one and CIs attaining the nominal level of coverage. When both models were misspecified, all three variance estimators tracked with the ESE, but bias was substantial, resulting in CIs with below nominal coverage.

Differences between variance estimators are apparent for scenarios where only one working model was correctly specified. When either the outcome model or the propensity model was misspecified, CIs based on the empirical sandwich variance estimator and the nonparametric bootstrap attained the nominal level of coverage, and estimated standard errors generally tracked closely with the ESE. In Figures \ref{fig:SimresVardiff} and \ref{fig:SimresVardiffNull}, note variance ratios for the empirical sandwich estimator and the nonparametric bootstrap are clustered around one for these scenarios. In contrast, the influence function based variance estimator was empirically biased when either working model was misspecified. As expected, it tended to overestimate the variance when the outcome model was misspecified, leading to SERs above one and conservative CI coverage. When the propensity model was misspecified but the outcome model was correctly specified, the influence function based variance estimator underestimated the variance, resulting in SERs below one. Bias in the influence function based variance estimator was more pronounced under the null. Under outcome model misspecification, SERs exceeded 2.0, resulting in approximately 100\% CI coverage. Under propensity misspecification, SERs were 0.89, resulting in CIs with below nominal coverage.

The empirical sandwich variance and nonparametric bootstrap estimators both demonstrated the expected doubly robust variance property, but there was a notable difference in the performance of the estimators. The nonparametric bootstrap generally demonstrated more variation than the empirical sandwich estimator, as evidenced by increased spread and more extreme outliers in Figures \ref{fig:SimresVardiff} and \ref{fig:SimresVardiffNull}. 

Additional simulations were conducted with $\sigma \in \{200,600\}$, with a binary outcome, and with $n=2000$. Twelve scenarios were considered. Details and results of the additional scenarios are provided in Appendix C. 

In summary, the simulation study demonstrated the theoretical properties explained in Section \ref{sec:methods}. That is, the empirical sandwich variance estimator and the nonparametric bootstrap were empirically unbiased when at least one of the two working models was correctly specified. The influence function based variance estimator is not consistent under misspecification of either working model. The influence function based variance estimator was conservative when the outcome model was misspecified and was generally anti-conservative when the propensity model was misspecified. The magnitude and direction of bias under misspecified working models varied across simulation scenarios.

\begin{table}
\caption{Simulation summary results, continuous outcome, $n=800$, $\sigma=400$, $5000$ simulations. Bias, ESE, SER, and 95\% CI coverage calculated for the ACE.}
\label{tab:simres}
\centering

\begin{tabular}{c l c c c c c c c c} 
\hline
  Scenario & Estimator & Bias & ESE & \makecell{SER, \\ ES} & \makecell{Cov, \\ ES (\%)} &  \makecell{SER, \\ NB} & \makecell{Cov, \\ NB (\%)} &  \makecell{SER, \\ IF} & \makecell{Cov, \\ IF (\%)}\\ 
  \hline 
\\
$ACE \approx -60$	& & & & & & & & & \\	
\\
CS	&	Classic	&	0.4	&	58.4	&	0.99	&	95	&	1.00	&	95	&	1.00	&	95	\\
	&	WR	&	0.4	&	58.3	&	0.99	&	95	&	1.00	&	95	&	0.99	&	95	\\
	&	TMLE	&	0.4	&	58.3	&	0.99	&	95	&	1.00	&	95	&	1.00	&	95	\\
MO	&	Classic	&	-0.4	&	60.0	&	0.99	&	95	&	1.02	&	96	&	1.07	&	97	\\
	&	WR	&	-1.6	&	59.2	&	0.99	&	95	&	1.00	&	95	&	1.05	&	96	\\
	&	TMLE	&	-0.5	&	59.7	&	0.99	&	95	&	1.01	&	95	&	1.06	&	96	\\
MP	&	Classic	&	0.3	&	57.8	&	1.00	&	95	&	1.00	&	95	&	0.97	&	94	\\
	&	WR	&	0.3	&	57.8	&	1.00	&	95	&	1.00	&	95	&	0.97	&	94	\\
	&	TMLE	&	0.3	&	57.8	&	1.00	&	95	&	1.00	&	95	&	0.97	&	94	\\
MB	&	Classic	&	-23.8	&	57.0	&	1.00	&	92	&	1.00	&	92	&	1.00	&	92	\\
	&	WR	&	-23.8	&	57.0	&	1.00	&	92	&	1.00	&	92	&	1.00	&	92	\\
	&	TMLE	&	-23.8	&	57.0	&	1.00	&	92	&	1.00	&	92	&	1.00	&	92	\\
$ACE = 0$		& & & & & & & & & \\
\\
CS	&	Classic	&	0.1	&	35.0	&	0.96	&	94	&	1.00	&	95	&	0.97	&	94	\\
	&	WR	&	0.1	&	34.8	&	0.96	&	94	&	0.98	&	94	&	0.95	&	94	\\
	&	TMLE	&	0.1	&	35.0	&	0.96	&	94	&	0.99	&	95	&	0.97	&	94	\\
MO	&	Classic	&	1.4	&	48.0	&	0.93	&	94	&	1.07	&	96	&	2.10	&	100	\\
	&	WR	&	2.5	&	44.8	&	0.94	&	93	&	1.02	&	95	&	2.18	&	100	\\
	&	TMLE	&	1.6	&	46.7	&	0.93	&	94	&	1.04	&	96	&	2.12	&	100	\\
MP	&	Classic	&	0.0	&	33.7	&	0.98	&	95	&	0.99	&	95	&	0.89	&	92	\\
	&	WR	&	-0.1	&	33.7	&	0.98	&	95	&	0.99	&	95	&	0.89	&	92	\\
	&	TMLE	&	0.0	&	33.7	&	0.98	&	95	&	0.99	&	95	&	0.89	&	92	\\
MB	&	Classic	&	158.8	&	75.9	&	0.99	&	44	&	0.99	&	44	&	0.99	&	44	\\
	&	WR	&	158.8	&	75.9	&	0.99	&	44	&	0.99	&	44	&	0.99	&	44	\\
	&	TMLE	&	158.8	&	75.9	&	0.99	&	44	&	0.99	&	44	&	0.99	&	44	\\

\hline
\end{tabular}

\begin{tablenotes}
      \item ESE=empirical standard error; SER=standard error ratio (ASE/ESE), where ASE=average estimated standard error; Cov = 95\% confidence interval coverage, i.e., the proportion of simulated samples for which the 95\% CI included ACE; ES=empirical sandwich variance estimator; NB=nonparametric bootstrap variance estimator, IF=influence function based variance estimator; CS=correct specification of both models; MO=misspecified outcome model, MP=misspecified propensity model; MB=misspecified both models; WR=weighted regression AIPW; Monte Carlo standard error for 95\% CI coverage was 0.3\% when coverage was 95\%. Results exclude 1 simulation where models did not converge. 
    \end{tablenotes}

\end{table}

\section{Discussion}
\label{sec:discussion}
Doubly robust estimators have gained popularity due to their ability to provide consistent point estimates when either an outcome or propensity model is correctly specified. Here, the commonly-used influence function based variance estimator is compared with the empirical sandwich variance estimator and the nonparametric bootstrap in conjunction with three doubly robust estimators: the classic AIPW estimator, the weighted regression AIPW estimator, and TMLE. For estimation of the average causal effect with observational data, the influence function based variance estimator is consistent only when both outcome and propensity models are correctly specified. In contrast, both the empirical sandwich variance estimator and the nonparametric bootstrap are doubly robust variance estimators. As such, confidence intervals constructed from these estimators are expected to provide nominal CI coverage when either model is correctly specified. 

This paper only considers variance estimation of the average causal effect with observational data. The influence function based variance estimator can be consistent under misspecification of one working model in some settings. For example, this variance estimator is consistent under outcome model misspecification in the randomized setting under additional assumptions \citep{williams2022optimising,van2024automated} and for estimation of some non-degenerate estimands \citep{haneuse2013estimation}.

While here consideration was limited to finite-dimensional parametric modeling approaches, machine learning approaches are commonly applied with doubly robust estimators. These approaches allow for more complex functional forms for continuous covariates than fully parametric approaches and include higher order interaction terms that may not be included in investigator-specified parametric models, often leading to estimators that are more robust to model misspecification \citep{zivich2014machine}. However, convergence of machine learning algorithms is typically slower and consistent variance estimation is more challenging than with parametric modeling approaches. Machine learning methods are typically not compatible with the estimating equations approach discussed in this paper, though alternative methods have been developed for doubly robust variance estimation in this context for TMLE estimators \citep{benkeser2017doubly}.

Use of the empirical sandwich variance estimator in conjunction with doubly robust estimators is not new, though its doubly robust property has not been emphasized. \cite{Lunceford2004} discuss use of the empirical sandwich variance estimator for estimating the variance of the classic AIPW estimator and note that it tends to be more stable than the influence function based variance estimator. However, based on the systematic review discussed in the introduction \citep{smith2023application}, the empirical sandwich variance estimator does not appear to be widely used with TMLE. While some existing software packages compute empirical sandwich variance estimators for doubly robust estimators (e.g., the ``causaltrt'' procedure in SAS and the ``dr'' procedure in Stata), other popular software packages compute the influence function based variance estimator but not the empirical sandwich variance estimator (e.g., the \texttt{AIPW} and \texttt{tmle} packages in R and the \texttt{zEpid} package in Python). We hope this work will allow for easier implementation of doubly robust point and variance estimation.

\section*{Supplementary Material} 
R and Python code for computing the different estimators along with the corresponding standard error estimators is available at https://github.com/bonnieshook/DoublyRobustVariance.

\section*{Bibliography}

	\bibliographystyle{apalike}
	\bibliography{bibliography}

\section*{Acknowledgements} 
This work was supported by the National Institutes of Health under award numbers R01 AI157758, K01 AI182506, R01 AI085073, P30 AI050410, and K01 AI177102. The content is solely the responsibility of the authors and does not necessarily represent the official views of the NIH. The authors thank Drs. Bradley Saul and Michael Hudgens at the University of North Carolina for coding support and helpful suggestions, respectively. \vspace*{-8pt}

	\clearpage
	
	\begin{appendices}
		\setcounter{equation}{0}
		\renewcommand{\theequation}{A.\arabic{equation}}

    \setcounter{table}{0}
    \renewcommand{\thetable}{A\arabic{table}}

    \setcounter{figure}{0}
    \renewcommand{\thefigure}{A\arabic{figure}}

\section*{Appendix A: parameter vectors and estimating functions for three doubly robust estimators with the M-estimation approach}

\label{s:webA_EE}

\smallskip
\noindent\textbf{Classic AIPW estimator}

For the classic AIPW estimator, $\theta=[\alpha, \gamma, \mu_{C}^1, \mu_{C}^0, DR_{C}]$. The estimating function $\psi_1$ is specified as
\[  \psi_1(O_i; \theta) = \begin{bmatrix}  \psi_\alpha \\ 
 \psi_\gamma  \\
 \psi_{\mu_{C1}}  \\
 \psi_{\mu_{C0}}  \\ 
 \psi_{C} 
\end{bmatrix} = \begin{bmatrix}  \{X_i-\mbox{expit}(G_{i}^T\alpha)\}G_{i} \\ 
 \{Y_i-\phi^{-1}(H_{i}^T\gamma)\}H_{i} \\
 \{X_iY_i-(X_i-e_i)a_1(Z_i,\gamma)\}e_i^{-1} - \mu_{C}^1\\
 \{(1-X_i)Y_i+(X_i-e_i)a_0(Z_i,\gamma)\}(1-e_i)^{-1} - \mu_{C}^0 \\
 \mu_{C}^1-\mu_{C}^0 - DR_{C} 
\end{bmatrix}\]
where $\psi_{\alpha}$ is the vector of score functions from the propensity score model, $\psi_{\gamma}$ is the vector of score functions for the outcome model (with link function $\phi$), and $H_i=h(X_i,Z_i)$ is a vector of predictors for the outcome model for participant $i$ for some user-specified function $h$ of $X_i$ and $Z_i$. The estimating functions $\psi_{\mu_{C1}}$ and $\psi_{\mu_{C0}}$ are for the causal means $\mu_{C}^1$ and $\mu_{C}^0$, respectively, and $\psi_{C}$ is a delta method transformation of the causal means for estimation of the $ACE$ using the classic AIPW, i.e., $\widehat{DR}_{C}$. \cite{shook2024exposure} demonstrate that this set of estimating equations is unbiased when at least one of the models is correctly specified.

\smallskip
\noindent\textbf{Weighted regression AIPW estimator}

For the weighted regression AIPW estimator, $\theta=[\alpha, \beta, \mu_{WR}^1,\mu_{WR}^0, DR_{WR}]$. The estimating function $\psi_2$ is specified as
\[  \psi_2(O_i; \theta) = \begin{bmatrix}  \psi_\alpha \\ 
 \psi_\beta  \\
 \psi_{\mu_{WR1}}  \\
 \psi_{\mu_{WR0}}  \\ 
 \psi_{WR} 
\end{bmatrix} = \begin{bmatrix}  \{X_i-\mbox{expit}(G_{i}^T\alpha)\}G_{i} \\ 
 W_i\{Y_i-\phi^{-1}(H_{i}^T\beta)\}H_{i} \\
  b_1(Z_i,\beta)- \mu_{WR}^1\\
  b_0(Z_i,\beta)- \mu_{WR}^0 \\
 \mu_{WR}^1-\mu_{WR}^0 - DR_{WR} 
\end{bmatrix}\]
where $\psi_{\beta}$ is the vector of IPTW-weighted score functions for the outcome model and $b_x(Z_i, \beta) = \phi^{-1}(H_i(x, Z_i))^\top  \beta$ for $x \in \{0,1\}$. The estimating functions $\psi_{\mu_{WR1}}$ and $\psi_{\mu_{WR0}}$ are for the causal means $\mu_{WR}^1$ and $\mu_{WR}^0$, respectively, and $\psi_{WR}$ is a delta method transformation of the causal means for estimation of the $ACE$ using the weighted regression AIPW estimator $\widehat{DR}_{WR}$. \cite{gabriel2023inverse} demonstrate unbiasedness of this set of estimating equations when at least one of the working models is correctly specified.

\smallskip
\noindent\textbf{TMLE}

For TMLE, $\theta=[\alpha, \gamma, \eta_1, \eta_0, \mu_{TMLE}^1, \mu_{TMLE}^0, DR_{TMLE}]$. For continuous outcomes, the estimating function $\psi_3$ is specified as
\[  \psi_3(O_i; \theta) = \begin{bmatrix}  \psi_\alpha \\ 
 \psi_\gamma  \\
 \psi_{t1} \\
 \psi_{t0} \\
 \psi_{\mu_{TMLE1}}  \\
 \psi_{\mu_{TMLE0}}  \\ 
 \psi_{TMLE}  
\end{bmatrix} = \begin{bmatrix}  \{X_i-\mbox{expit}(G_{i}^T\alpha)\}G_{i} \\ 
 \{Y^*_i-\phi^{-1}(H_{i}^T\gamma)\}H_{i} \\
 X_ie_i^{-1}(Y^*_i-\mbox{expit}[\eta_1+\mbox{logit}\{a_1(Z_i,\gamma)\}]) \\
 (1-X_i)(1-e_i)^{-1}(Y^*_i-\mbox{expit}[\eta_0+\mbox{logit}\{a_0(Z_i,\gamma)\}]) \\ 
  c_1(O_i, \gamma, \eta_1) - \mu_{TMLE}^1\\
  c_0(O_i, \gamma, \eta_0) - \mu_{TMLE}^0 \\
 \mu_{TMLE}^1-\mu_{TMLE}^0 - DR_{TMLE} 
\end{bmatrix}\]
where $\psi_{\gamma}$ is the vector of score functions for the scaled outcome model. The estimating functions $\psi_{t1}$ and $\psi_{t0}$ are for the two targeting models, while $\psi_{\mu_{TMLE1}}$ and $\psi_{\mu_{TMLE0}}$ are for the causal means $\mu_{TMLE}^1$ and $\mu_{TMLE}^0$, respectively. Note $c_x(O_i, \gamma, \eta_x)=\mbox{expit}[\mbox{logit}\{a_x(Z_i,\gamma)\}+\eta_x](b-a)+a$ are the pseudo-outcomes following the targeting step, as outlined in Section 2.2.3 of the main text. Finally, $\psi_{TMLE}$ is a delta method transformation of the causal means for estimation of the $ACE$ using $\widehat{DR}_{TMLE}$. Unbiasedness of $\psi_3(O_i; \theta)$ when at least one working model is correctly specified is demonstrated in Appendix B.

\section*{Appendix B: Consistency proof of TMLE estimator under correct specification of either the outcome or propensity model}

\label{s:webB_TechDeriv}

Recall the TMLE estimating function $\psi_3$ is specified as
\[  \psi_3(O_i; \theta) = \begin{bmatrix}  \psi_\alpha \\ 
 \psi_\gamma  \\
 \psi_{t1} \\
 \psi_{t0} \\
 \psi_{\mu_{TMLE1}}  \\
 \psi_{\mu_{TMLE0}}  \\ 
 \psi_{TMLE}  
\end{bmatrix} = \begin{bmatrix}  \{X_i-\mbox{expit}(G_{i}^T\alpha)\}G_{i} \\ 
 \{Y^*_i-\phi^{-1}(H_{i}^T\gamma)\}H_{i} \\
 X_ie_i^{-1}(Y^*_i-\mbox{expit}[\eta_1+\mbox{logit}\{a_1(Z_i,\gamma)\}]) \\
 (1-X_i)(1-e_i)^{-1}(Y^*_i-\mbox{expit}[\eta_0+\mbox{logit}\{a_0(Z_i,\gamma)\}]) \\ 
 c_1(O_i, \gamma, \eta_1) - \mu_{TMLE}^1\\
 c_0(O_i, \gamma, \eta_0) - \mu_{TMLE}^0 \\
 \mu_{TMLE}^1-\mu_{TMLE}^0 - DR_{TMLE} 
\end{bmatrix}\] where $\theta=[\alpha, \gamma , \eta_1 , \eta_0 , \mu_{TMLE}^1 , \mu_{TMLE}^0 , DR_{TMLE}]$ is the vector of parameters.
Let

\noindent{$\tilde{\theta}
= [\tilde{\alpha} , \tilde{\gamma} , \tilde{\eta}_{1} , \tilde{\eta}_{0} , \tilde{\mu}_{TMLE}^1 , \tilde{\mu}_{TMLE}^0 , \widetilde{DR}_{TMLE}]$} be the solution of 
$E\{\psi_3(O_i; \theta)\} = 0$ and $\widehat{DR}_{TMLE}$ be the proposed TMLE estimator of ACE, obtained by solving the estimating equation $n^{-1} \sum_i \psi_3(O_i;\theta) = 0$. Then, if either the outcome model or the propensity score model is correctly specified, $\sqrt{n} (\widehat{DR}_{TMLE} - ACE) \stackrel{d}{\longrightarrow} N(0,\sigma^2(\tilde{\theta}))$, where $\sigma^2(\tilde{\theta})$ is the bottom right element of $V({\tilde{\theta}}) = V({\theta}) |_{\theta = \tilde{\theta}}$, and
$V({\theta}) = A(\theta)^{-1}B(\theta)A(\theta)^{-T}$,
with
$A(\theta) = E\{-\partial \psi_3(O_i; \theta) / \partial \theta \}$,
$B(\theta) = E\{\psi_3(O_i; \theta) \psi_3(O_i; \theta)^\top \}$,
which is a doubly robust variance estimator of $\widehat{DR}_{TMLE}$.

\noindent
\textit{Proof.}

\noindent{Let} $\hat{\theta}$ be the solution of the estimating equation $n^{-1} \sum_i \psi_3(O_i;\theta) = 0$.
Under suitable regularity conditions \citep{Stefanski2002}, $\sqrt{n} (\hat{\theta} - \tilde{\theta})
	\stackrel{d}{\longrightarrow} N(0,V(\tilde{\theta}))$. Note the final element of $\hat{\theta}$ is $\widehat{DR}_{TMLE}$ and therefore
$\sqrt{n} (\widehat{DR}_{TMLE} - \widetilde{DR}_{TMLE})	\stackrel{d}{\longrightarrow} N(0,\sigma^2(\tilde{\theta}))$. It remains to show that if either the outcome model or the propensity score model is correctly specified, $\widetilde{DR}_{TMLE} = ACE$.

First, let 
$\theta_0
= [\alpha_0, \gamma_0, \eta_{10}, \eta_{00}, \mu_{TMLE0}^1, \mu_{TMLE0}^0, DR_{TMLE0}]
$
be the true parameter values from the data generating process.
Then,
$P(X_i = 1 | Z_i) = e(Z_i, \alpha_0) = \mbox{expit}(G_i^T \alpha_0)$
and
$E(Y_i^* | Z_i, X_i) = \phi^{-1}(H_i^T \gamma_0) 
= X_i a_1(Z_i, \gamma_0) + (1-X_i) a_0(Z_i, \gamma_0)$.

Now assume the outcome model is correctly specified, 
i.e., $\tilde{\gamma} = \gamma_0$ and
$E(Y_i^* | Z_i, X_i)
= X_i a_1(Z_i, \gamma_0) + (1-X_i) a_0(Z_i, \gamma_0)$.
From $E\{\psi_{t1}(O_i; \tilde{\theta})\} = 0$,
\begin{align*}
	0 
	=&
	E\{\psi_{t1}(O_i; \tilde{\theta})\} 
	\\
	=&
	E[E\{\psi_{t1}(O_i; \tilde{\theta}) | X_i\}]
	\\
	=&
	E[E\{
		X_i 
		e(Z_i, \tilde{\alpha})^{-1} 
		(Y_i^* - \mbox{expit}[\tilde{\eta}_1 + \mbox{logit}\{a_1(Z_i, \gamma_0)\}]) 
	| X_i\}]
	\\
	=&
	E\{
		e(Z_i, \tilde{\alpha})^{-1} 
		(Y_i^* - \mbox{expit}[\tilde{\eta}_1 + \mbox{logit}\{a_1(Z_i, \gamma_0)\}]) 
	| X_i = 1\} P(X_i = 1)
	\\
	=&
	E[
		E\{
			e(Z_i, \tilde{\alpha})^{-1} 
			(Y_i^* - \mbox{expit}[\tilde{\eta}_1 + \mbox{logit}\{a_1(Z_i, \gamma_0)\}]) 
			| Z_i, X_i = 1
		\}
		| X_i = 1
	] 
	P(X_i = 1)
	\\
	=&
	E[
		e(Z_i, \tilde{\alpha})^{-1}
		\{
			E(Y_i^* | Z_i, X_i = 1)
			-
			\mbox{expit}[\tilde{\eta}_1 + \mbox{logit}\{a_1(Z_i, \gamma_0)\}]
		\}
		| X_i = 1
	] 
	P(X_i = 1)
	\\
	=&
	E[
		e(Z_i, \tilde{\alpha})^{-1}
		\{
			a_1(Z_i, \gamma_0)
			-
			\mbox{expit}[\tilde{\eta}_1 + \mbox{logit}\{a_1(Z_i, \gamma_0)\}]
		\}
		| X_i = 1
	] 
	P(X_i = 1)
	\\
	=&
	E[
		e(Z_i, \tilde{\alpha})^{-1}
		F(\tilde{\eta}_1)
		| X_i = 1
	] 
	P(X_i = 1)
\end{align*}
where
$F(t) = \mbox{expit}(d) - \mbox{expit}(t + d)$ and $d = \mbox{logit}\{a_1(Z_i, \gamma_0)\}$.
Since the function 
$F(t)$
is strictly decreasing in $t$,
it can be shown that the last line in the above is strictly decreasing in $\tilde{\eta}_1$ and attains 0 only at $\tilde{\eta}_1 = 0$.
Therefore, $\tilde{\eta}_1 = 0$.
Then, from $E\{\psi_{\mu_{TMLE1}}(O_i; \tilde{\theta})\} = 0$,
\begin{align*}
	\tilde{\mu}_{TMLE}^1 
	=& 
	E(Y_{i, TMLE}^1)
	\\
	=&
	E\{
		\mbox{expit}[\tilde{\eta}_1 + \mbox{logit}\{a_1(Z_i, \gamma_0)\}](b-a) + a
	\}
	\\
	=&
	E\{a_1(Z_i, \gamma_0)\}(b-a) + a
	\\
	=&
	E[
		E\{
			Y_i^* | Z_i, X_i = 1
		\}
	]
	(b-a) + a
	\\
	=&
	E[
		E\{
			(Y_i-a)/(b-a) | Z_i, X_i = 1
		\}
	]
	(b-a) + a
	\\
	=&
	E\{E(Y_i| Z_i, X_i = 1)\}
	\\
	=&
	\mu^1.
\end{align*}
Similarly, $\tilde{\mu}_{TMLE}^0  = \mu^0$.

Next, assume the propensity score model is correctly specified, i.e., $\tilde{\alpha} = \alpha_0$
and
$P(X_i=1|Z_i) = e(Z_i, \alpha_0)$.
From $E\{\psi_{t1}(O_i; \tilde{\theta})\} = 0$,
\begin{align*}
	0 
	=&
	E\{\psi_{t1}(O_i; \tilde{\theta})\} 
	\\
	=&
	E[E\{\psi_{t1}(O_i; \tilde{\theta}) | Z_i\}]
	\\
	=&
	E[
		E\{
			X_i 
			e(Z_i, \alpha_0)^{-1} 
			\{(Y_i^1-a)/(b-a) 
			- 
			\mbox{expit}[\tilde{\eta}_1 + \mbox{logit}\{a_1(Z_i, \tilde{\gamma})\}]) \}
			| Z_i
		\}
	]
	\\
	=&
	E(
		E(X_i | Z_i)
		e(Z_i, \alpha_0)^{-1} 
		[
			E\{(Y_i^1-a)/(b-a)|Z_i\}
			- 
			\mbox{expit}[\tilde{\eta}_1 + \mbox{logit}\{a_1(Z_i, \tilde{\gamma})\}]
		]
	)
	\\
	=&
	E[E\{(Y_i^1-a)/(b-a)|Z_i\}]
	-
	E(\mbox{expit}[\tilde{\eta}_1 + \mbox{logit}\{a_1(Z_i, \tilde{\gamma})\}])
	\\
	=&
	(\mu^1-a)/(b-a)
	-
	E(\mbox{expit}[\tilde{\eta}_1 + \mbox{logit}\{a_1(Z_i, \tilde{\gamma})\}]).
\end{align*}
Therefore,
$E(\mbox{expit}[\tilde{\eta}_1 + \mbox{logit}\{a_1(Z_i, \tilde{\gamma})\}]) = (\mu^1-a)/(b-a)$.
Then, from $E\{\psi_{\mu_{TMLE1}}(O_i; \tilde{\theta})\} = 0$,
\begin{align*}
	\tilde{\mu}_{TMLE}^1 
	=& 
	E\{c_x(O_i, \gamma, \eta_x)\}
	\\
	=&
	E\{
		\mbox{expit}[\tilde{\eta}_1 + \mbox{logit}\{a_1(Z_i, \tilde{\gamma})\}](b-a) + a
	\}
	\\
	=&
	\mu^1.
\end{align*}
Similarly, $\tilde{\mu}_{TMLE}^0  = \mu^0$.

In conclusion,
if either the outcome model or the propensity score model is correctly specified, 
$\widetilde{DR}_{TMLE} = \tilde{\mu}_{TMLE}^1  - \tilde{\mu}_{TMLE}^0  =  \mu^1 - \mu^0 = ACE$
which proves the stated result.

\section*{Appendix C: Additional Simulation Scenarios}
\label{s:webC_sims}

\subsection*{Additional Simulation Setup}
In addition to simulations described in Section 4 of the main text, 10 additional simulation scenarios were considered:
\begin{enumerate}
    \item continuous outcome, $n=800$, $\sigma=200$, $ACE \approx -60$
    \item continuous outcome, $n=800$, $\sigma=600$, $ACE \approx -60$
    \item continuous outcome, $n=2000$, $\sigma=200$, $ACE \approx -60$
    \item continuous outcome, $n=2000$, $\sigma=400$, $ACE \approx -60$
    \item continuous outcome, $n=2000$, $\sigma=600$, $ACE \approx -60$
    \item continuous outcome, $n=2000$, $\sigma=400$, $ACE=0$ (null)
    \item binary outcome, $n=800$, $ACE \approx 0.045$
    \item binary outcome, $n=800$, $ACE=0$ (null)
    \item binary outcome, $n=2000$, $ACE \approx 0.045$
    \item binary outcome, $n=2000$, $ACE=0$ (null)
\end{enumerate}

The continuous outcome scenarios followed the same data generating mechanism described in Section 4 of the main text, with the specified sample size and value of $\sigma$. For binary outcome scenarios, the distributions of $Z_1$, $Z_2$, and $Z_3$ were the same as the continuous outcome scenarios, but the exposure and outcomes differed. The exposure $X$ was simulated from a Bernoulli distribution with mean $\mbox{expit}(16-0.1 Z_1 + 2.5 Z_2 - Z_3 - 0.02 Z_1 Z_2 + 0.005 Z_1 Z_3)$. Potential outcomes $Y^0$ and $Y^1$ under no exposure and  exposure, respectively, were simulated from Bernoulli distributions with means $\mbox{expit}\{-2+0.5(Z_1-155)+0.75Z_2-0.8Z_2(Z_1-155)-0.2(Z_1-155)(1-x)-0.1Z_2(1-x)+Z_2(Z_1-155)(1-x)\}$ for $x \in \{0,1\}$. Under this data generating mechanism, $ACE \approx 0.045$. To examine performance of the estimators under the null, simulations were also conducted for the binary outcome where $E(Y^1)$ was equal to $E(Y^0)$, as defined above, such that $ACE=0$.

\subsection*{Additional Simulation Results}
Results of the additional simulation scenarios are presented in Appendix Tables A2-A11 and Appendix Figures A2-A11. When $n=2000$, empirical bias in point estimates, empirical sandwich variance estimates, and nonparametric bootstrap estimates under correct specification of at least one model was further reduced. However, an increase in the sample size did not mitigate bias in the influence function based variance estimator when one model was misspecified. For continuous outcomes, under misspecification of the outcome model, the conservative property of the influence function based variance estimator decreased as $\sigma$ increased. Under misspecification of the propensity model, influence function based variance estimators became increasingly anti-conservative as $\sigma$ increased.

For the binary outcome, the conservative property of the influence function based variance estimator under outcome model misspecification was more pronounced than for the continuous outcome. Notably, under misspecified propensity models, the influence function based variance estimator was slightly conservative when there was non-null effect but slightly anti-conservative under the null. These findings highlight that the direction and magnitude of bias under propensity model misspecification vary across data generating mechanisms.

\begin{table} [h]
\caption{Baseline characteristics of IPOP participants by anemia status}
\label{tab:App_Covs}
\centering
\begin{tabular}{l l c c} 
\hline
 &  & \makecell{No anemia \\ $n=664$} & \makecell{Anemia \\ $n=118$}  \\
\hline
\\
Age category	&	18-24	&	156 (23\%)	&	28 (24\%)	\\
	&	25-29	&	190 (29\%)	&	33 (28\%)	\\
	&	30-34	&	192 (29\%)	&	31 (26\%)	\\
	&	35-39	&	100 (15\%)	&	21 (18\%)	\\
	&	40+	&	26 (4\%)	&	5 (4\%)	\\
Number of prior births	&	0	&	125 (19\%)	&	30 (25\%)	\\
	&	1	&	160 (24\%)	&	25 (21\%)	\\
	&	2	&	157 (24\%)	&	28 (24\%)	\\
	&	3	&	125 (19\%)	&	18 (15\%)	\\
	&	4+	&	97 (15\%)	&	17 (14\%)	\\
First trimester	&		&	120 (18\%)	&	12 (10\%)	\\
ART use	&		&	648 (98\%)	&	109 (92\%)	\\
Alcohol use during pregnancy	&		&	64 (10\%)	&	9 (8\%)	\\
Height (cm)	&	Median (Q1,Q3)	&	155 (150, 160)	&	154 (150, 159)	\\
	&	Mean (SD)	&	155 (8)	&	153 (8)	\\
	&	Min, Max	&	131, 185	&	132, 170	\\

  \hline 
\end{tabular}
\begin{tablenotes}
      \item Note: Anemia defined as baseline hemoglobin below 10.5 g/dL 

    \end{tablenotes}
\end{table}

\begin{table}
\caption{Simulation summary results, continuous outcome, $n=800$, $\sigma=200$, $5000$ simulations. Bias, ESE, SER, and 95\% CI coverage calculated for the ACE.}
\centering

\begin{tabular}{c l c c c c c c c c} 
\hline
  Scenario & Estimator & Bias & ESE & \makecell{SER, \\ ES} & \makecell{Cov, \\ ES (\%)} &  \makecell{SER, \\ NB} & \makecell{Cov, \\ NB (\%)} &  \makecell{SER, \\ IF} & \makecell{Cov, \\ IF (\%)}\\ 
  \hline 
CS	&	Classic	&	0.4	&	50.1	&	1.00	&	95	&	1.00	&	95	&	1.00	&	95	\\
	&	WR	&	0.4	&	50.1	&	1.00	&	95	&	1.00	&	95	&	1.00	&	95	\\
	&	TMLE	&	0.3	&	50.1	&	1.00	&	95	&	1.00	&	95	&	1.00	&	95	\\
MO	&	Classic	&	-0.3	&	51.9	&	0.99	&	95	&	1.02	&	96	&	1.10	&	97	\\
	&	WR	&	-1.6	&	51.3	&	0.99	&	95	&	1.00	&	95	&	1.07	&	96	\\
	&	TMLE	&	-0.4	&	51.6	&	0.99	&	95	&	1.01	&	95	&	1.08	&	96	\\
MP	&	Classic	&	0.4	&	50.0	&	1.00	&	95	&	1.00	&	95	&	0.99	&	95	\\
	&	WR	&	0.3	&	50.0	&	1.00	&	95	&	1.00	&	95	&	0.99	&	95	\\
	&	TMLE	&	0.3	&	50.0	&	1.00	&	95	&	1.00	&	95	&	0.99	&	95	\\
MB	&	Classic	&	-23.5	&	50.7	&	0.99	&	92	&	1.00	&	92	&	0.99	&	92	\\
	&	WR	&	-23.5	&	50.7	&	0.99	&	92	&	1.00	&	92	&	0.99	&	92	\\
	&	TMLE	&	-23.5	&	50.7	&	0.99	&	92	&	1.00	&	92	&	0.99	&	92	\\

\hline
\end{tabular}

\begin{tablenotes}
      \item ESE=empirical standard error; SER=standard error ratio (ASE/ESE), where ASE=average estimated standard error; Cov = 95\% confidence interval coverage; ES=empirical sandwich variance estimator; NB=nonparametric bootstrap variance estimator, IF=influence function based variance estimator; CS=correct specification of both models; MO=misspecified outcome model, MP=misspecified propensity model; MB=misspecified both models; WR=weighted regression AIPW; $ACE$ was approximately $-60$; Monte Carlo standard error for 95\% CI coverage was 0.3\% when coverage was 95\%. Results exclude 15 simulations where models failed to converge.
    \end{tablenotes}

\end{table}

\begin{table}
\caption{Simulation summary results, continuous outcome, $n=800$, $\sigma=600$, $5000$ simulations. Bias, ESE, SER, and 95\% CI coverage calculated for the ACE.}
\centering

\begin{tabular}{c l c c c c c c c c} 
\hline
  Scenario & Estimator & Bias & ESE & \makecell{SER, \\ ES} & \makecell{Cov, \\ ES (\%)} &  \makecell{SER, \\ NB} & \makecell{Cov, \\ NB (\%)} &  \makecell{SER, \\ IF} & \makecell{Cov, \\ IF (\%)}\\ 
  \hline 
CS	&	Classic	&	0.4	&	70.1	&	0.99	&	94	&	1.00	&	95	&	0.99	&	95	\\
	&	WR	&	0.4	&	69.9	&	0.98	&	94	&	1.00	&	95	&	0.98	&	94	\\
	&	TMLE	&	0.4	&	70.1	&	0.99	&	94	&	1.00	&	95	&	0.99	&	95	\\
MO	&	Classic	&	-0.4	&	71.5	&	0.98	&	95	&	1.01	&	95	&	1.05	&	96	\\
	&	WR	&	-1.7	&	70.6	&	0.98	&	95	&	1.00	&	95	&	1.03	&	96	\\
	&	TMLE	&	-0.5	&	71.2	&	0.98	&	95	&	1.00	&	95	&	1.04	&	95	\\
MP	&	Classic	&	0.3	&	68.9	&	0.99	&	95	&	1.00	&	95	&	0.95	&	93	\\
	&	WR	&	0.2	&	68.9	&	0.99	&	95	&	1.00	&	95	&	0.95	&	93	\\
	&	TMLE	&	0.3	&	68.9	&	0.99	&	95	&	1.00	&	95	&	0.95	&	93	\\
MB	&	Classic	&	-24.1	&	66.2	&	1.00	&	93	&	1.00	&	93	&	1.00	&	93	\\
	&	WR	&	-24.1	&	66.2	&	1.00	&	93	&	1.00	&	93	&	1.00	&	93	\\
	&	TMLE	&	-24.1	&	66.2	&	1.00	&	93	&	1.00	&	93	&	1.00	&	93	\\

\hline
\end{tabular}

\begin{tablenotes}
      \item ESE=empirical standard error; SER=standard error ratio (ASE/ESE), where ASE=average estimated standard error; Cov = 95\% confidence interval coverage; ES=empirical sandwich variance estimator; NB=nonparametric bootstrap variance estimator, IF=influence function based variance estimator; CS=correct specification of both models; MO=misspecified outcome model, MP=misspecified propensity model; MB=misspecified both models; WR=weighted regression AIPW; $ACE$ was approximately $-60$; Monte Carlo standard error for 95\% CI coverage was 0.3\% when coverage was 95\%. 
    \end{tablenotes}

\end{table}

\begin{table}
\caption{Simulation summary results, continuous outcome, $n=2000$, $\sigma=200$, $5000$ simulations. Bias, ESE, SER, and 95\% CI coverage calculated for the ACE.}
\centering

\begin{tabular}{c l c c c c c c c c} 
\hline
  Scenario & Estimator & Bias & ESE & \makecell{SER, \\ ES} & \makecell{Cov, \\ ES (\%)} &  \makecell{SER, \\ NB} & \makecell{Cov, \\ NB (\%)} &  \makecell{SER, \\ IF} & \makecell{Cov, \\ IF (\%)}\\ 
  \hline 
CS	&	Classic	&	0.0	&	31.6	&	1.00	&	95	&	1.00	&	95	&	1.00	&	95	\\
	&	WR	&	0.0	&	31.5	&	1.00	&	95	&	1.00	&	95	&	1.00	&	95	\\
	&	TMLE	&	0.0	&	31.6	&	1.00	&	95	&	1.00	&	95	&	1.00	&	95	\\
MO	&	Classic	&	-0.2	&	32.3	&	1.00	&	95	&	1.01	&	95	&	1.11	&	97	\\
	&	WR	&	-0.8	&	32.1	&	1.00	&	95	&	1.01	&	95	&	1.09	&	97	\\
	&	TMLE	&	-0.2	&	32.1	&	1.00	&	95	&	1.01	&	95	&	1.09	&	97	\\
MP	&	Classic	&	0.0	&	31.4	&	1.00	&	95	&	1.00	&	95	&	0.99	&	95	\\
	&	WR	&	0.0	&	31.4	&	1.00	&	95	&	1.00	&	95	&	0.99	&	95	\\
	&	TMLE	&	0.0	&	31.4	&	1.00	&	95	&	1.00	&	95	&	0.99	&	95	\\
MB	&	Classic	&	-23.6	&	32.0	&	1.00	&	89	&	1.00	&	89	&	1.00	&	89	\\
	&	WR	&	-23.6	&	32.0	&	1.00	&	89	&	1.00	&	89	&	1.00	&	89	\\
	&	TMLE	&	-23.6	&	32.0	&	1.00	&	89	&	1.00	&	89	&	1.00	&	89	\\

\hline
\end{tabular}

\begin{tablenotes}
      \item ESE=empirical standard error; SER=standard error ratio (ASE/ESE), where ASE=average estimated standard error; Cov = 95\% confidence interval coverage; ES=empirical sandwich variance estimator; NB=nonparametric bootstrap variance estimator, IF=influence function based variance estimator; CS=correct specification of both models; MO=misspecified outcome model, MP=misspecified propensity model; MB=misspecified both models; WR=weighted regression AIPW; $ACE$ was approximately $-60$; Monte Carlo standard error for 95\% CI coverage was 0.3\% when coverage was 95\%. 
    \end{tablenotes}

\end{table}

\begin{table}
\caption{Simulation summary results, continuous outcome, $n=2000$, $\sigma=400$, $5000$ simulations. Bias, ESE, SER, and 95\% CI coverage calculated for the ACE.}
\centering

\begin{tabular}{c l c c c c c c c c} 
\hline
  Scenario & Estimator & Bias & ESE & \makecell{SER, \\ ES} & \makecell{Cov, \\ ES (\%)} &  \makecell{SER, \\ NB} & \makecell{Cov, \\ NB (\%)} &  \makecell{SER, \\ IF} & \makecell{Cov, \\ IF (\%)}\\ 
  \hline 
CS	&	Classic	&	-0.1	&	36.6	&	1.00	&	95	&	1.01	&	95	&	1.00	&	95	\\
	&	WR	&	-0.1	&	36.6	&	1.00	&	95	&	1.00	&	95	&	1.00	&	95	\\
	&	TMLE	&	-0.1	&	36.6	&	1.00	&	95	&	1.01	&	95	&	1.00	&	95	\\
MO	&	Classic	&	-0.3	&	37.2	&	1.01	&	95	&	1.01	&	95	&	1.09	&	97	\\
	&	WR	&	-0.9	&	37.0	&	1.00	&	95	&	1.01	&	95	&	1.07	&	96	\\
	&	TMLE	&	-0.4	&	37.1	&	1.01	&	95	&	1.01	&	95	&	1.07	&	96	\\
MP	&	Classic	&	-0.2	&	36.2	&	1.01	&	95	&	1.01	&	95	&	0.98	&	94	\\
	&	WR	&	-0.2	&	36.2	&	1.01	&	95	&	1.01	&	95	&	0.98	&	95	\\
	&	TMLE	&	-0.2	&	36.2	&	1.01	&	95	&	1.01	&	95	&	0.98	&	94	\\
MB	&	Classic	&	-23.8	&	35.9	&	1.00	&	90	&	1.00	&	90	&	1.00	&	90	\\
	&	WR	&	-23.8	&	35.9	&	1.00	&	90	&	1.00	&	90	&	1.00	&	90	\\
	&	TMLE	&	-23.8	&	35.9	&	1.00	&	90	&	1.00	&	90	&	1.00	&	90	\\

\hline
\end{tabular}

\begin{tablenotes}
      \item ESE=empirical standard error; SER=standard error ratio (ASE/ESE), where ASE=average estimated standard error; Cov = 95\% confidence interval coverage; ES=empirical sandwich variance estimator; NB=nonparametric bootstrap variance estimator, IF=influence function based variance estimator; CS=correct specification of both models; MO=misspecified outcome model, MP=misspecified propensity model; MB=misspecified both models; WR=weighted regression AIPW; $ACE$ was approximately $-60$; Monte Carlo standard error for 95\% CI coverage was 0.3\% when coverage was 95\%. 
    \end{tablenotes}

\end{table}

\begin{table}
\caption{Simulation summary results, continuous outcome, $n=2000$, $\sigma=600$, $5000$ simulations. Bias, ESE, SER, and 95\% CI coverage calculated for the ACE.}
\centering

\begin{tabular}{c l c c c c c c c c} 
\hline
  Scenario & Estimator & Bias & ESE & \makecell{SER, \\ ES} & \makecell{Cov, \\ ES (\%)} &  \makecell{SER, \\ NB} & \makecell{Cov, \\ NB (\%)} &  \makecell{SER, \\ IF} & \makecell{Cov, \\ IF (\%)}\\ 
  \hline 
CS	&	Classic	&	-0.2	&	43.8	&	1.00	&	95	&	1.01	&	95	&	1.00	&	95	\\
	&	WR	&	-0.2	&	43.8	&	1.00	&	95	&	1.00	&	95	&	1.00	&	95	\\
	&	TMLE	&	-0.2	&	43.8	&	1.00	&	95	&	1.01	&	95	&	1.00	&	95	\\
MO	&	Classic	&	-0.5	&	44.3	&	1.01	&	95	&	1.01	&	96	&	1.07	&	96	\\
	&	WR	&	-1.1	&	44.0	&	1.00	&	95	&	1.01	&	95	&	1.05	&	96	\\
	&	TMLE	&	-0.5	&	44.2	&	1.00	&	95	&	1.01	&	95	&	1.05	&	96	\\
MP	&	Classic	&	-0.4	&	43.1	&	1.01	&	95	&	1.01	&	95	&	0.96	&	94	\\
	&	WR	&	-0.4	&	43.1	&	1.01	&	95	&	1.01	&	95	&	0.96	&	94	\\
	&	TMLE	&	-0.4	&	43.1	&	1.01	&	95	&	1.01	&	95	&	0.96	&	94	\\
MB	&	Classic	&	-24.0	&	41.6	&	1.00	&	91	&	1.00	&	91	&	1.00	&	91	\\
	&	WR	&	-24.0	&	41.6	&	1.00	&	91	&	1.00	&	91	&	1.00	&	91	\\
	&	TMLE	&	-24.0	&	41.6	&	1.00	&	91	&	1.00	&	91	&	1.00	&	91	\\

\hline
\end{tabular}

\begin{tablenotes}
      \item ESE=empirical standard error; SER=standard error ratio (ASE/ESE), where ASE=average estimated standard error; Cov = 95\% confidence interval coverage; ES=empirical sandwich variance estimator; NB=nonparametric bootstrap variance estimator, IF=influence function based variance estimator; CS=correct specification of both models; MO=misspecified outcome model, MP=misspecified propensity model; MB=misspecified both models; WR=weighted regression AIPW; $ACE$ was approximately $-60$; Monte Carlo standard error for 95\% CI coverage was 0.3\% when coverage was 95\%. 
    \end{tablenotes}

\end{table}

\begin{table}
\caption{Simulation summary results, continuous outcome, $n=2000$, $\sigma=400$, $5000$ simulations under the null. Bias, ESE, SER, and 95\% CI coverage calculated for the ACE.}
\centering

\begin{tabular}{c l c c c c c c c c} 
\hline
  Scenario & Estimator & Bias & ESE & \makecell{SER, \\ ES} & \makecell{Cov, \\ ES (\%)} &  \makecell{SER, \\ NB} & \makecell{Cov, \\ NB (\%)} &  \makecell{SER, \\ IF} & \makecell{Cov, \\ IF (\%)}\\ 
  \hline 
CS	&	Classic	&	-0.2	&	21.6	&	1.00	&	95	&	1.00	&	95	&	1.00	&	95	\\
	&	WR	&	-0.2	&	21.6	&	0.99	&	95	&	1.00	&	95	&	0.99	&	95	\\
	&	TMLE	&	-0.2	&	21.6	&	0.99	&	95	&	1.00	&	95	&	1.00	&	95	\\
MO	&	Classic	&	0.3	&	28.9	&	0.96	&	95	&	1.01	&	96	&	2.20	&	100	\\
	&	WR	&	0.9	&	27.3	&	0.97	&	94	&	1.00	&	95	&	2.26	&	100	\\
	&	TMLE	&	0.4	&	28.3	&	0.97	&	95	&	1.01	&	95	&	2.21	&	100	\\
MP	&	Classic	&	-0.3	&	21.0	&	1.00	&	95	&	1.01	&	95	&	0.91	&	93	\\
	&	WR	&	-0.3	&	21.0	&	1.00	&	95	&	1.01	&	95	&	0.91	&	93	\\
	&	TMLE	&	-0.3	&	21.0	&	1.00	&	95	&	1.00	&	95	&	0.91	&	93	\\
MB	&	Classic	&	159.4	&	47.5	&	1.00	&	8	&	1.00	&	9	&	1.00	&	8	\\
	&	WR	&	159.4	&	47.5	&	1.00	&	8	&	1.00	&	9	&	1.00	&	8	\\
	&	TMLE	&	159.4	&	47.5	&	1.00	&	8	&	1.00	&	9	&	1.00	&	8	\\

\hline
\end{tabular}

\begin{tablenotes}
      \item ESE=empirical standard error; SER=standard error ratio (ASE/ESE), where ASE=average estimated standard error; Cov = 95\% confidence interval coverage; ES=empirical sandwich variance estimator; NB=nonparametric bootstrap variance estimator, IF=influence function based variance estimator; CS=correct specification of both models; MO=misspecified outcome model, MP=misspecified propensity model; MB=misspecified both models; WR=weighted regression AIPW; $ACE$ was $0$; Monte Carlo standard error for 95\% CI coverage was 0.3\% when coverage was 95\%. 
    \end{tablenotes}

\end{table}

\begin{table}
\caption{Simulation summary results, binary outcome, $n=800$,  $5000$ simulations. Bias, ESE, SER, and 95\% CI coverage calculated for the ACE.}
\centering

\begin{tabular}{c l c c c c c c c c} 
\hline
  Scenario & Estimator & Bias & ESE & \makecell{SER, \\ ES} & \makecell{Cov, \\ ES (\%)} &  \makecell{SER, \\ NB} & \makecell{Cov, \\ NB (\%)} &  \makecell{SER, \\ IF} & \makecell{Cov, \\ IF (\%)}\\ 
  \hline 

CS	&	Classic	&	0.0	&	2.6	&	0.99	&	95	&	0.99	&	95	&	0.99	&	95	\\
	&	WR	&	0.0	&	2.6	&	0.99	&	95	&	0.99	&	95	&	0.99	&	95	\\
	&	TMLE	&	0.0	&	2.6	&	0.99	&	95	&	0.99	&	95	&	0.99	&	95	\\
MO	&	Classic	&	0.0	&	2.9	&	0.97	&	94	&	1.00	&	95	&	1.28	&	99	\\
	&	WR	&	-0.1	&	2.8	&	0.97	&	94	&	1.07	&	95	&	1.24	&	99	\\
	&	TMLE	&	0.0	&	2.9	&	0.97	&	94	&	0.99	&	95	&	1.22	&	98	\\
MP	&	Classic	&	0.0	&	2.6	&	0.99	&	95	&	0.99	&	95	&	1.02	&	95	\\
	&	WR	&	0.0	&	2.6	&	0.99	&	95	&	0.99	&	95	&	1.02	&	95	\\
	&	TMLE	&	0.0	&	2.6	&	0.99	&	95	&	0.99	&	95	&	1.02	&	95	\\
MB	&	Classic	&	-15.7	&	3.2	&	0.99	&	0	&	0.99	&	0	&	0.99	&	0	\\
	&	WR	&	-15.7	&	3.2	&	0.99	&	0	&	0.99	&	0	&	0.99	&	0	\\
	&	TMLE	&	-15.7	&	3.2	&	0.99	&	0	&	0.99	&	0	&	0.99	&	0	\\

\hline
\end{tabular}

\begin{tablenotes}
      \item ESE=empirical standard error; SER=standard error ratio (ASE/ESE), where ASE=average estimated standard error; Cov = 95\% confidence interval coverage; ES=empirical sandwich variance estimator; NB=nonparametric bootstrap variance estimator, IF=influence function based variance estimator; CS=correct specification of both models; MO=misspecified outcome model, MP=misspecified propensity model; MB=misspecified both models; WR=weighted regression AIPW; $ACE$ was approximately $0.045$; Monte Carlo standard error for 95\% CI coverage was 0.3\% when coverage was 95\%. Results exclude eight simulations where models failed to converge. 
    \end{tablenotes}

\end{table}

\begin{table}
\caption{Simulation summary results, binary outcome, $n=800$,  $5000$ simulations under the null. Bias, ESE, SER, and 95\% CI coverage calculated for the ACE.}
\centering

\begin{tabular}{c l c c c c c c c c} 
\hline
  Scenario & Estimator & Bias & ESE & \makecell{SER, \\ ES} & \makecell{Cov, \\ ES (\%)} &  \makecell{SER, \\ NB} & \makecell{Cov, \\ NB (\%)} &  \makecell{SER, \\ IF} & \makecell{Cov, \\ IF (\%)}\\ 
  \hline 

CS	&	Classic	&	0.0	&	2.4	&	0.99	&	95	&	0.99	&	95	&	0.99	&	95	\\
	&	WR	&	0.0	&	2.4	&	0.99	&	95	&	0.99	&	95	&	0.99	&	95	\\
	&	TMLE	&	0.0	&	2.4	&	0.99	&	95	&	1.06	&	95	&	0.99	&	95	\\
MO	&	Classic	&	0.0	&	2.5	&	1.00	&	96	&	1.03	&	96	&	1.49	&	100	\\
	&	WR	&	-0.2	&	2.5	&	1.00	&	95	&	1.04	&	96	&	1.39	&	99	\\
	&	TMLE	&	0.0	&	2.5	&	1.00	&	95	&	1.02	&	96	&	1.39	&	100	\\
MP	&	Classic	&	0.0	&	2.4	&	0.99	&	95	&	1.00	&	95	&	0.98	&	94	\\
	&	WR	&	0.0	&	2.4	&	0.99	&	95	&	1.00	&	95	&	0.98	&	95	\\
	&	TMLE	&	0.0	&	2.4	&	0.99	&	95	&	1.06	&	95	&	0.98	&	95	\\
MB	&	Classic	&	-20.1	&	3.1	&	0.99	&	0	&	0.99	&	0	&	0.99	&	0	\\
	&	WR	&	-20.1	&	3.1	&	0.99	&	0	&	0.99	&	0	&	0.99	&	0	\\
	&	TMLE	&	-20.1	&	3.1	&	0.99	&	0	&	0.99	&	0	&	0.99	&	0	\\

\hline
\end{tabular}

\begin{tablenotes}
      \item ESE=empirical standard error; SER=standard error ratio (ASE/ESE), where ASE=average estimated standard error; Cov = 95\% confidence interval coverage; ES=empirical sandwich variance estimator; NB=nonparametric bootstrap variance estimator, IF=influence function based variance estimator; CS=correct specification of both models; MO=misspecified outcome model, MP=misspecified propensity model; MB=misspecified both models; WR=weighted regression AIPW; $ACE$ was $0$; Monte Carlo standard error for 95\% CI coverage was 0.3\% when coverage was 95\%. Results exclude eight simulations where models failed to converge. 
    \end{tablenotes}

\end{table}

\begin{table}
\caption{Simulation summary results, binary outcome, $n=2000$,  $5000$ simulations. Bias, ESE, SER, and 95\% CI coverage calculated for the ACE.}
\centering
\begin{tabular}{c l c c c c c c c c} 
\hline
  Scenario & Estimator & Bias & ESE & \makecell{SER, \\ ES} & \makecell{Cov, \\ ES (\%)} &  \makecell{SER, \\ NB} & \makecell{Cov, \\ NB (\%)} &  \makecell{SER, \\ IF} & \makecell{Cov, \\ IF (\%)}\\ 
  \hline 
CS	&	Classic	&	0.0	&	1.6	&	0.99	&	95	&	1.00	&	95	&	0.99	&	95	\\
	&	WR	&	0.0	&	1.6	&	0.99	&	95	&	1.00	&	95	&	0.99	&	95	\\
	&	TMLE	&	0.0	&	1.6	&	0.99	&	95	&	1.00	&	95	&	0.99	&	95	\\
MO	&	Classic	&	0.0	&	1.8	&	1.00	&	95	&	1.01	&	95	&	1.30	&	99	\\
	&	WR	&	-0.1	&	1.7	&	0.99	&	95	&	1.10	&	95	&	1.26	&	99	\\
	&	TMLE	&	0.0	&	1.8	&	0.99	&	95	&	1.00	&	95	&	1.25	&	99	\\
MP	&	Classic	&	0.0	&	1.6	&	0.99	&	95	&	0.99	&	95	&	1.03	&	96	\\
	&	WR	&	0.0	&	1.6	&	0.99	&	95	&	0.99	&	95	&	1.03	&	96	\\
	&	TMLE	&	0.0	&	1.6	&	0.99	&	95	&	0.99	&	95	&	1.03	&	96	\\
MB	&	Classic	&	-15.6	&	2.0	&	1.00	&	0	&	1.00	&	0	&	1.00	&	0	\\
	&	WR	&	-15.6	&	2.0	&	1.00	&	0	&	1.00	&	0	&	1.00	&	0	\\
	&	TMLE	&	-15.6	&	2.0	&	1.00	&	0	&	1.00	&	0	&	1.00	&	0	\\

\hline
\end{tabular}

\begin{tablenotes}
      \item ESE=empirical standard error; SER=standard error ratio (ASE/ESE), where ASE=average estimated standard error; Cov = 95\% confidence interval coverage; ES=empirical sandwich variance estimator; NB=nonparametric bootstrap variance estimator, IF=influence function based variance estimator; CS=correct specification of both models; MO=misspecified outcome model, MP=misspecified propensity model; MB=misspecified both models; WR=weighted regression AIPW; $ACE$ was approximately $0.045$; Monte Carlo standard error for 95\% CI coverage was 0.3\% when coverage was 95\%. Results exclude four simulations where models failed to converge. 
    \end{tablenotes}

\end{table}

\begin{table}
\caption{Simulation summary results, binary outcome, $n=2000$,  $5000$ simulations under the null. Bias, ESE, SER, and 95\% CI coverage calculated for the ACE.}
\centering

\begin{tabular}{c l c c c c c c c c} 
\hline
  Scenario & Estimator & Bias & ESE & \makecell{SER, \\ ES} & \makecell{Cov, \\ ES (\%)} &  \makecell{SER, \\ NB} & \makecell{Cov, \\ NB (\%)} &  \makecell{SER, \\ IF} & \makecell{Cov, \\ IF (\%)}\\ 
  \hline 
CS	&	Classic	&	0.0	&	1.5	&	0.98	&	94	&	0.98	&	94	&	0.98	&	94	\\
	&	WR	&	0.0	&	1.5	&	0.98	&	94	&	0.98	&	95	&	0.98	&	94	\\
	&	TMLE	&	0.0	&	1.5	&	0.98	&	94	&	0.98	&	95	&	0.98	&	94	\\
MO	&	Classic	&	0.0	&	1.6	&	0.99	&	95	&	1.00	&	95	&	1.49	&	100	\\
	&	WR	&	-0.1	&	1.6	&	0.99	&	95	&	1.03	&	95	&	1.39	&	99	\\
	&	TMLE	&	0.0	&	1.6	&	0.99	&	95	&	1.00	&	95	&	1.39	&	99	\\
MP	&	Classic	&	0.0	&	1.5	&	0.98	&	95	&	0.98	&	95	&	0.97	&	94	\\
	&	WR	&	0.0	&	1.5	&	0.98	&	95	&	0.98	&	95	&	0.97	&	94	\\
	&	TMLE	&	0.0	&	1.5	&	0.98	&	95	&	0.98	&	95	&	0.97	&	94	\\
MB	&	Classic	&	-20.0	&	2.0	&	0.98	&	0	&	0.98	&	0	&	0.98	&	0	\\
	&	WR	&	-20.0	&	2.0	&	0.98	&	0	&	0.98	&	0	&	0.98	&	0	\\
	&	TMLE	&	-20.0	&	2.0	&	0.98	&	0	&	0.98	&	0	&	0.98	&	0	\\

\hline
\end{tabular}

\begin{tablenotes}
      \item ESE=empirical standard error; SER=standard error ratio (ASE/ESE), where ASE=average estimated standard error; Cov = 95\% confidence interval coverage; ES=empirical sandwich variance estimator; NB=nonparametric bootstrap variance estimator, IF=influence function based variance estimator; CS=correct specification of both models; MO=misspecified outcome model, MP=misspecified propensity model; MB=misspecified both models; WR=weighted regression AIPW; $ACE$ was $0$; Monte Carlo standard error for 95\% CI coverage was 0.3\% when coverage was 95\%. Results exclude one simulation where models failed to converge. 
    \end{tablenotes}

\end{table}

\begin{figure}
\begin{center}
  \includegraphics[width=0.5\columnwidth]{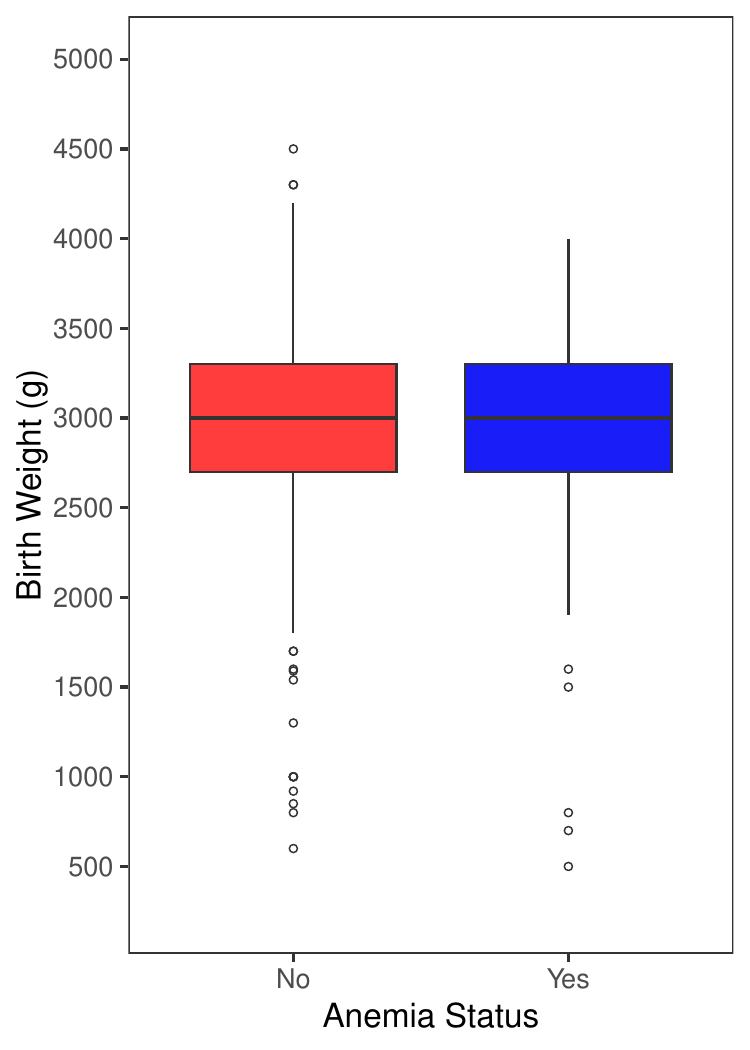}
  \caption{Boxplot of birth weights among $n=782$ IPOP participants by anemia status.}
\end{center}
\end{figure}

\begin{figure}
\begin{center}
  \includegraphics[width=\columnwidth]{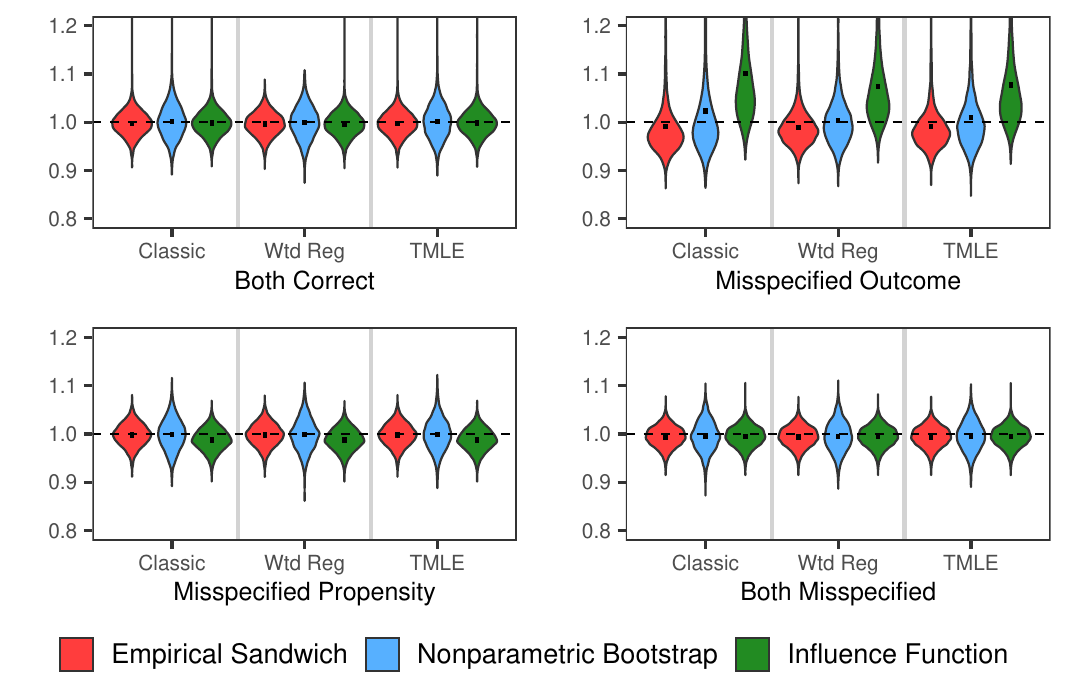}
  \caption{Ratio between each simulation's estimated standard error and the empirical standard error by estimator and model specification, continuous outcome, n=800, $\sigma=200$, 5000 simulations. $ACE$ was approximately -60. Black squares denote the mean variance ratio (=SER). Results exclude 15 simulations where models failed to converge. The $0.14\%$ of correct model specification simulations, $4.5\%$ of misspecified outcome model simulations, and $0.06\%$ of misspecified propensity model simulations where the ratio was above 1.2 or below 0.8 are not displayed. }
\end{center}
\end{figure}

\begin{figure}
\begin{center}
  \includegraphics[width=\columnwidth]{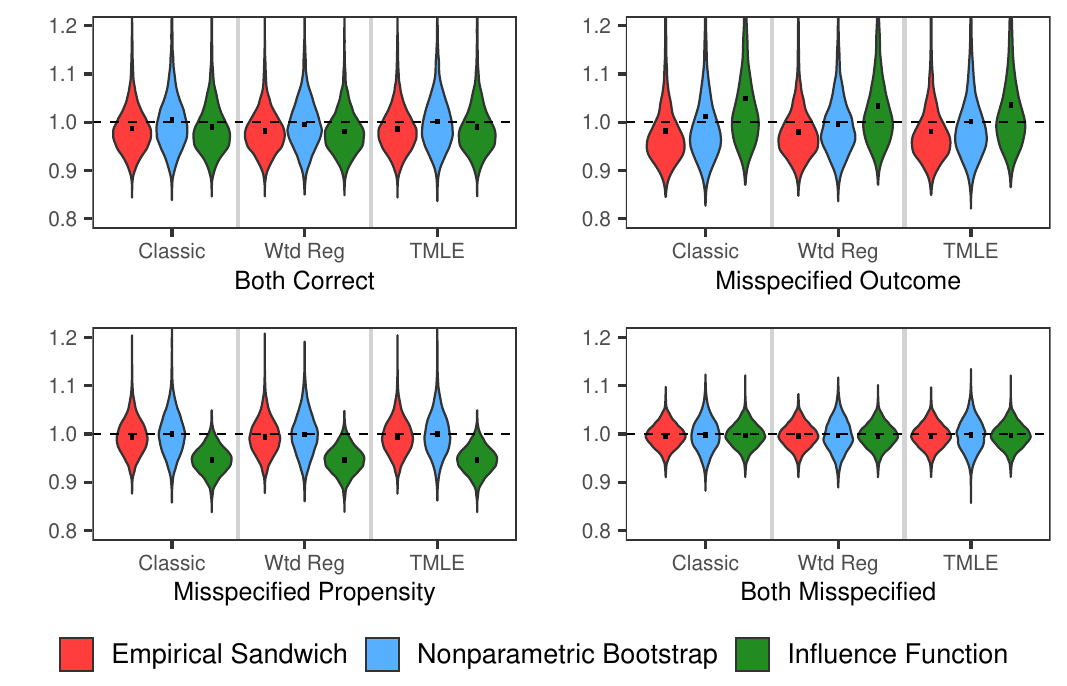}
  \caption{Ratio between each simulation's estimated standard error and the empirical standard error by estimator and model specification, continuous outcome, n=800, $\sigma=600$, 5000 simulations. $ACE$ was approximately -60. Black squares denote the mean variance ratio (=SER). The $0.77\%$ of correct model specification simulations, $3.66\%$ of misspecified outcome model simulations, and $0.01\%$ of misspecified propensity model simulations where the ratio was above 1.2 or below 0.8 are not displayed.}
\end{center}
\end{figure}

\begin{figure}
\begin{center}
  \includegraphics[width=\columnwidth]{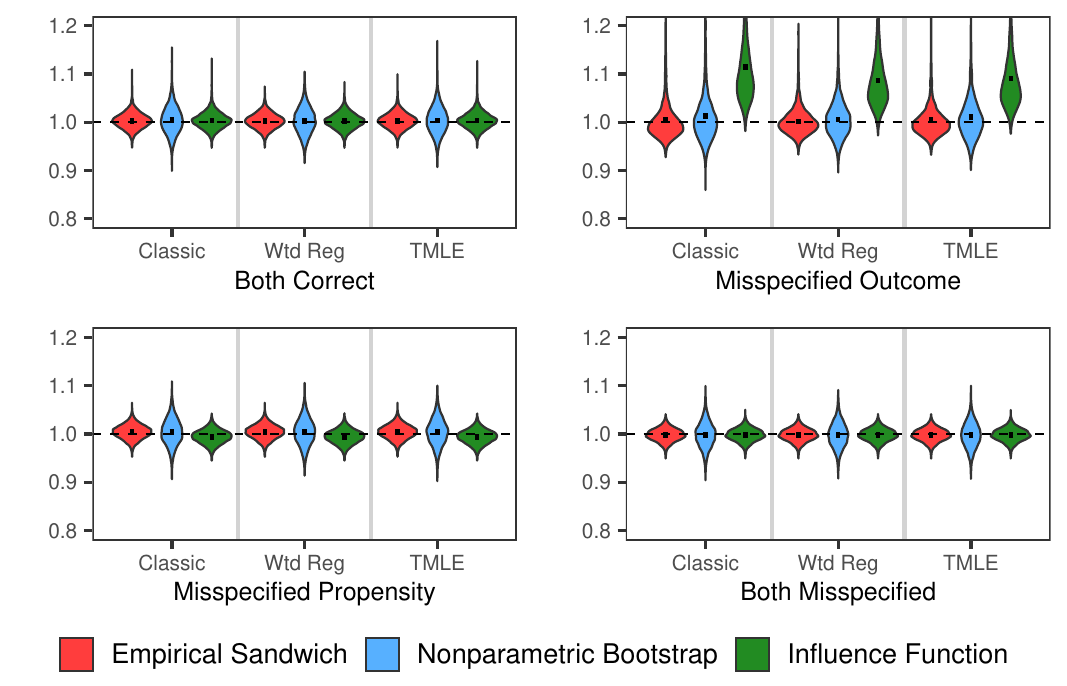}
  \caption{Ratio between each simulation's estimated standard error and the empirical standard error by estimator and model specification, continuous outcome, n=2000, $\sigma=200$, 5000 simulations. $ACE$ was approximately -60. Black squares denote the mean variance ratio (=SER). The $2.9\%$ of misspecified outcome model simulations where the ratio was above 1.2 or below 0.8 are not displayed.}
\end{center}
\end{figure}

\begin{figure}
\begin{center}
  \includegraphics[width=\columnwidth]{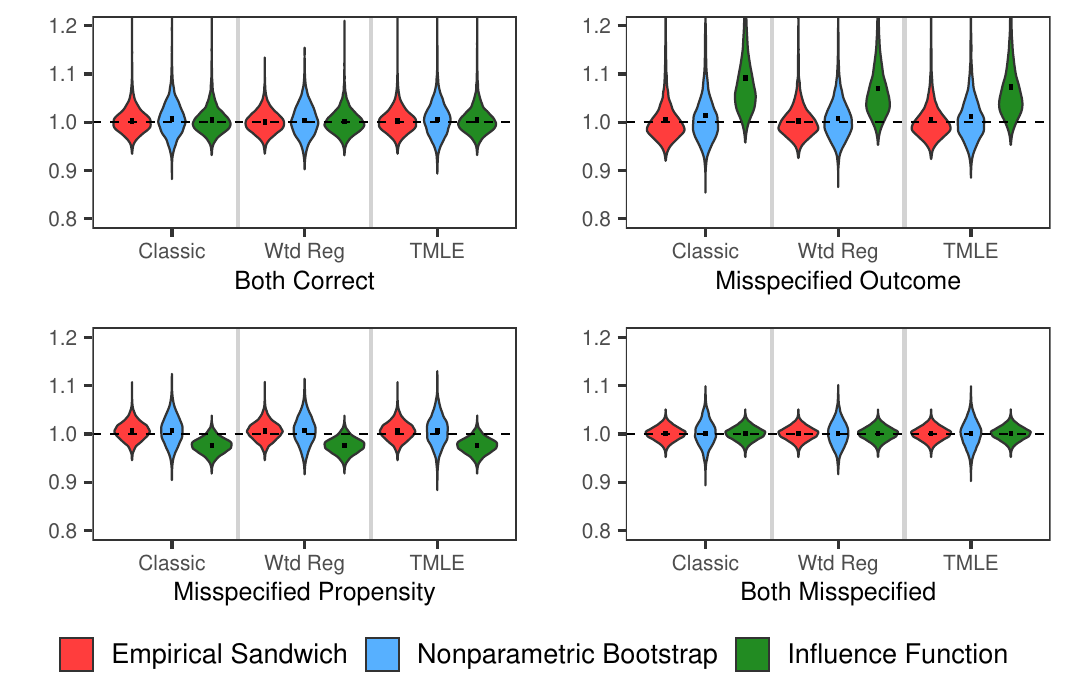}
  \caption{Ratio between each simulation's estimated standard error and the empirical standard error by estimator and model specification, continuous outcome, n=2000, $\sigma=400$, 5000 simulations. $ACE$ was approximately -60. Black squares denote the mean variance ratio (=SER). The $0.08\%$ of correct model specification simulations and $2.61\%$ of misspecified outcome model simulations where the ratio was above 1.2 or below 0.8 are not displayed.}
\end{center}
\end{figure}

\begin{figure}
\begin{center}
  \includegraphics[width=\columnwidth]{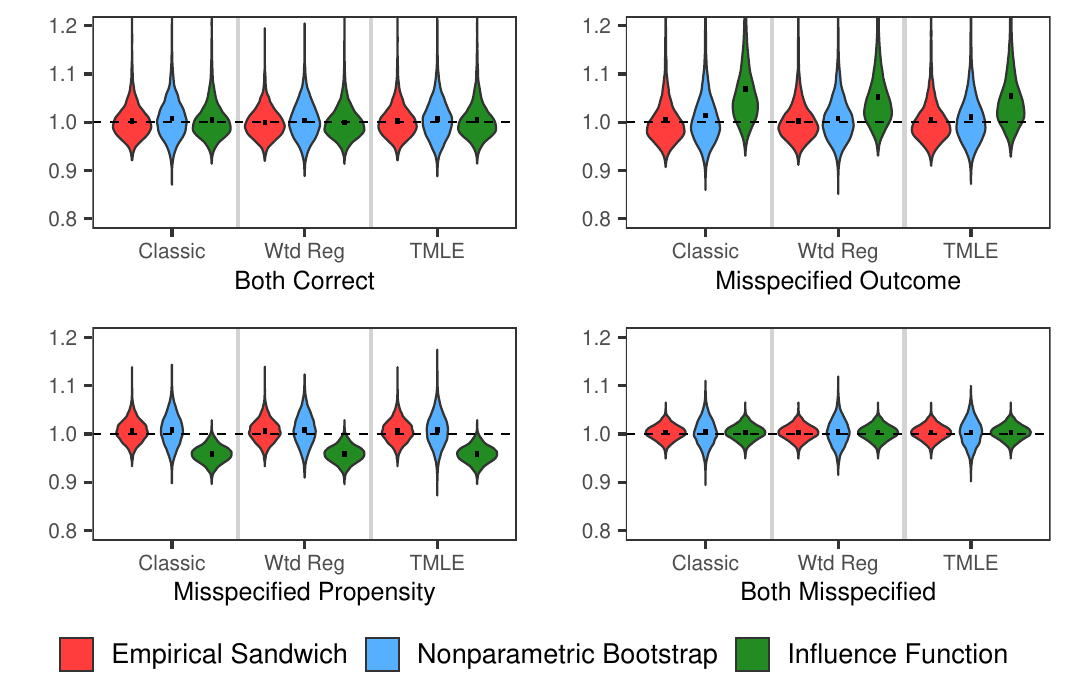}
  \caption{Ratio between each simulation's estimated standard error and the empirical standard error by estimator and model specification, continuous outcome, n=2000, $\sigma=600$, 5000 simulations. $ACE$ was approximately -60. Black squares denote the mean variance ratio (=SER). The $0.19\%$ of correct model specification simulations and $2.33\%$ of misspecified outcome model simulations where the ratio was above 1.2 or below 0.8 are not displayed.}
\end{center}
\end{figure}

\begin{figure}
\begin{center}
  \includegraphics[width=\columnwidth]{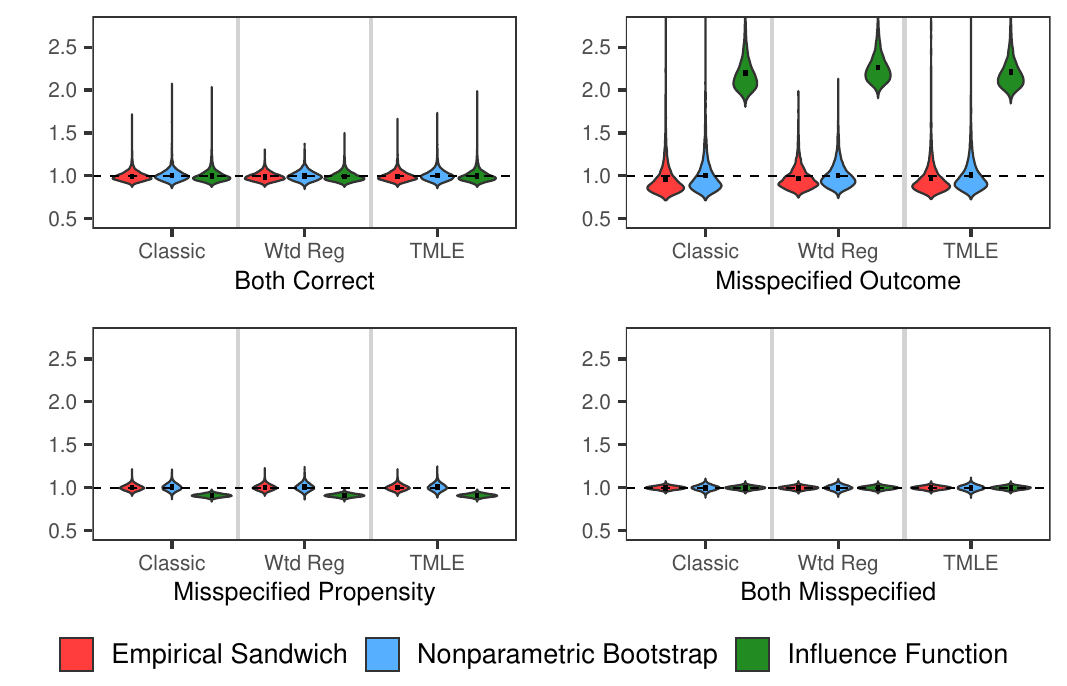}
  \caption{Ratio between each simulation's estimated standard error and the empirical standard error by estimator and model specification, continuous outcome, n=2000, $\sigma=400$, 5000 simulations under the null. Black squares denote the mean variance ratio (=SER). The $0.90\%$ of misspecified outcome model specification simulations where the ratio was above 2.75 or below 0.5 are not displayed.}
\end{center}
\end{figure}

\begin{figure}
\begin{center}
  \includegraphics[width=\columnwidth]{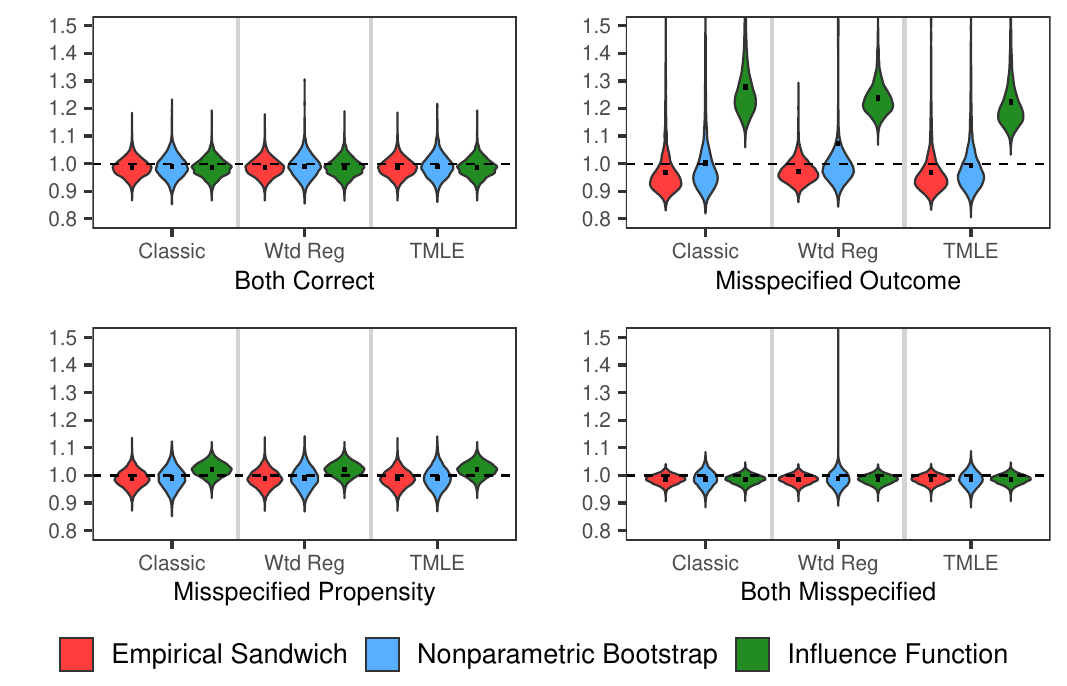}
  \caption{Ratio between each simulation's estimated standard error and the empirical standard error by estimator and model specification, binary outcome, n=800, 5000 simulations. $ACE$ was approximately 0.045. Black squares denote the mean variance ratio (=SER). Results exclude eight simulations where models failed to converge. The $1.63\%$ of misspecified outcome model simulations and $0.01\%$ of simulations where both models were misspecified where the ratio was above 1.5 or below 0.8 are not displayed.}
\end{center}
\end{figure}

\begin{figure}
\begin{center}
  \includegraphics[width=\columnwidth]{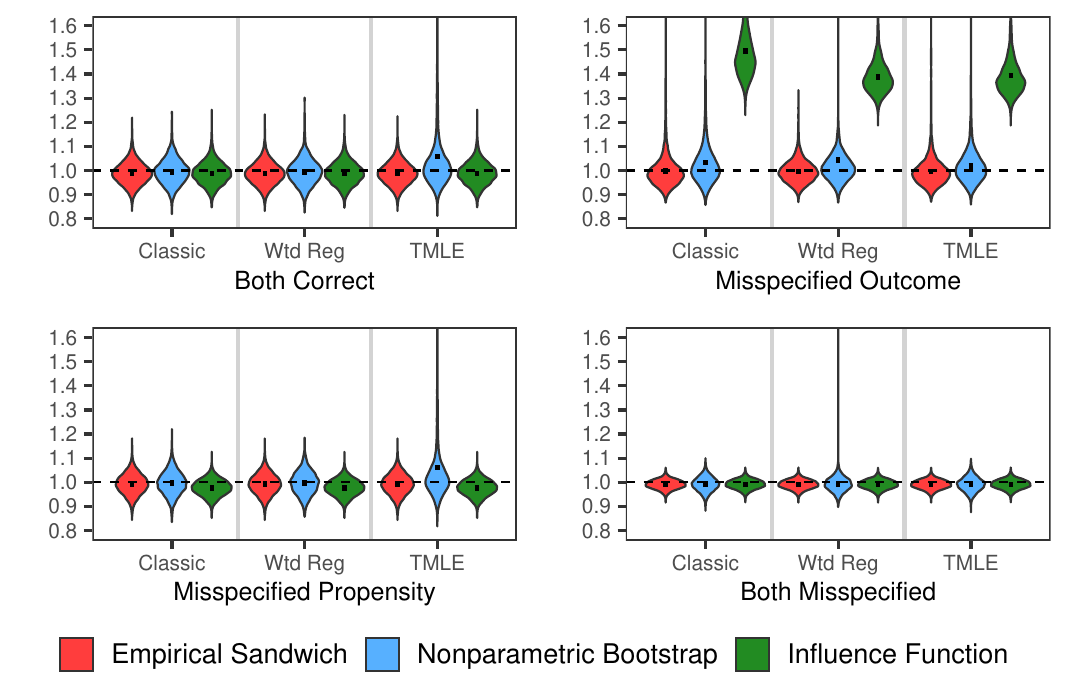}
  \caption{Ratio between each simulation's estimated standard error and the empirical standard error by estimator and model specification, binary outcome, n=800, 5000 simulations under the null. Black squares denote the mean variance ratio (=SER). Results exclude eight simulations where models failed to converge. The $0.27\%$ of correct model specification simulations, $2.26\%$ of misspecified outcome model simulations, $0.29\%$ of misspecified propensity simulations, and $0.01\%$ of simulations where both models were misspecified where the ratio was above 1.6 or below 0.8 are not displayed.}
\end{center}
\end{figure}

\begin{figure}
\begin{center}
  \includegraphics[width=\columnwidth]{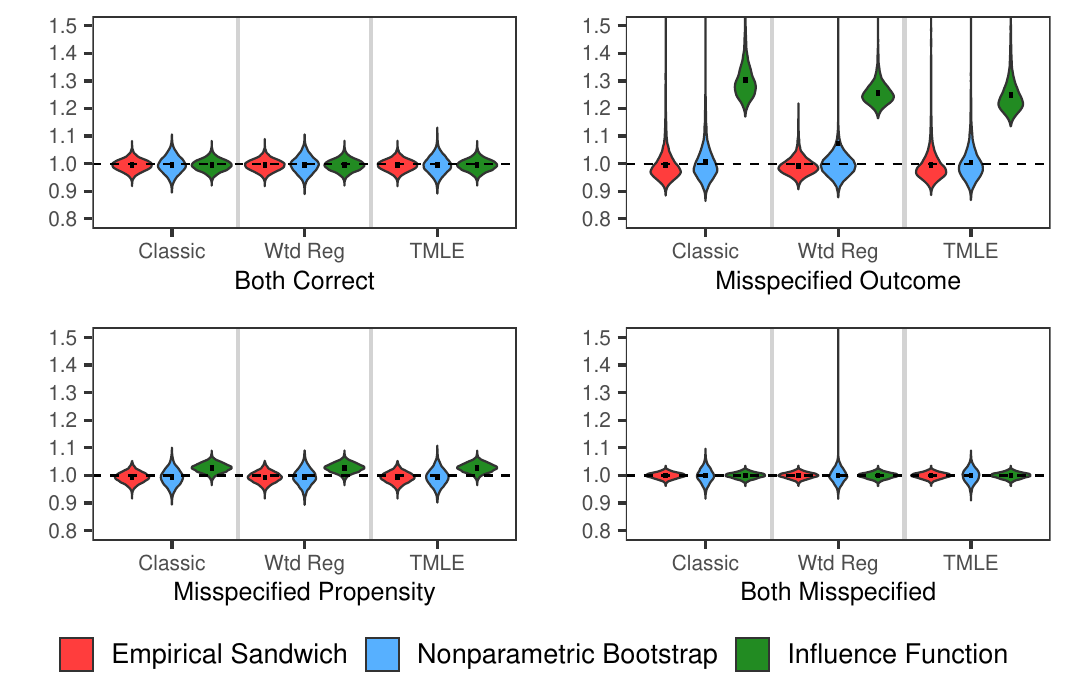}
  \caption{Ratio between each simulation's estimated standard error and the empirical standard error by estimator and model specification, binary outcome, n=2000, 5000 simulations. $ACE$ was approximately 0.045. Black squares denote the mean variance ratio (=SER). Results exclude four simulations where models failed to converge. The $0.72\%$ of misspecified outcome model simulations and $0.002\%$ of simulations where both models were misspecified where the ratio was above 1.5 or below 0.8 are not displayed.}
\end{center}
\end{figure}

\begin{figure}
\begin{center}
  \includegraphics[width=\columnwidth]{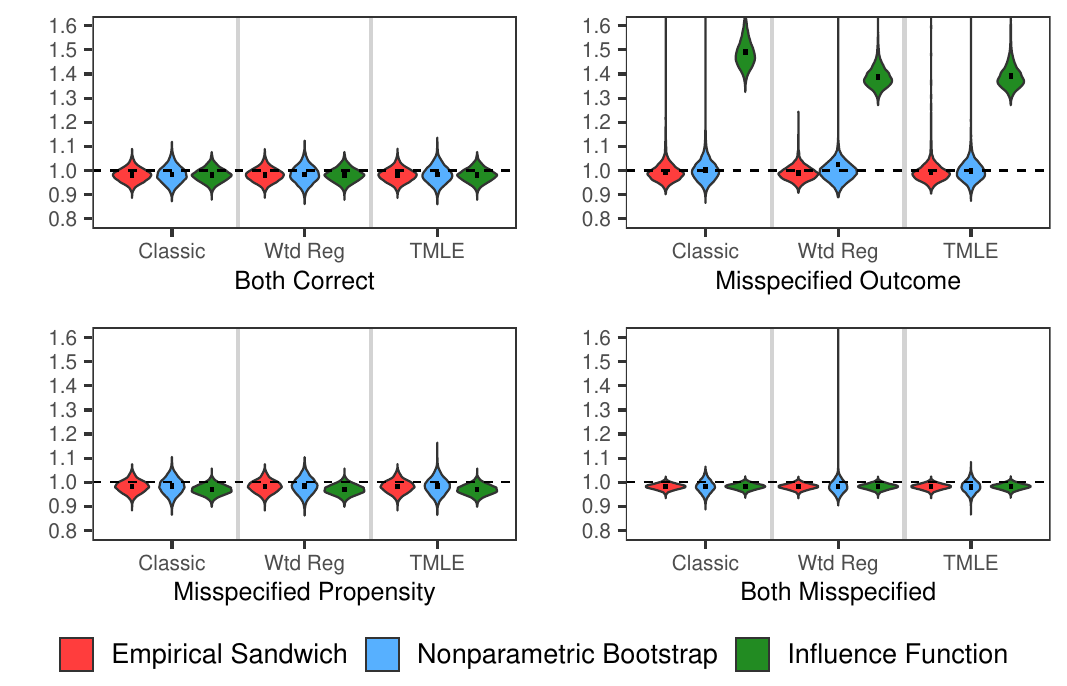}
  \caption{Ratio between each simulation's estimated standard error and the empirical standard error by estimator and model specification, binary outcome, n=2000, 5000 simulations under the null. Black squares denote the mean variance ratio (=SER). Results exclude one simulation where models failed to converge. The  $0.86\%$ of misspecified outcome model simulations and $0.002\%$ of simulations where both models were misspecified where the ratio was above 1.6 or below 0.8 are not displayed.}
\end{center}
\end{figure}

	\end{appendices}	
	
	\label{lastpage}
	
\end{document}